\documentclass[aps,prc,reprint,eqsecnum,showpacs,floatfix]{revtex4-1}

\usepackage{graphicx,epstopdf}
\usepackage[T1]{fontenc}
\usepackage[utf8]{inputenc}
\usepackage{hyperref}
\usepackage{color}
\usepackage{subfigure}
\usepackage{amsmath}
\usepackage{amssymb}
\usepackage{epsfig}
\usepackage{amsfonts}
\usepackage{mathtools}
\usepackage{bbm}
\usepackage{bm}
\usepackage{float}
\usepackage{xcolor}

\def\rvec{{\bf r}}

\def\kvec{{\bf k}}
\def\pvec{{\bf p}}
\def\qvec{{\bf q}}

\def\bra#1{\left\langle#1\right|}
\def\ket#1{\left|#1\right\rangle}

\def\he#1{$^{#1}$He}
\def\half{\frac{1}{2}}

\def\Im{{\cal I}m}
\def\I{{\rm i}}

\def\1{\mathbbm{1}}
\def\bra#1{\bigl\langle{ #1} \bigr|}
\def\ket#1{\bigl|{ #1} \bigr\rangle}

\def\ovlp#1#2{\bigl\langle{ #1}\big|{#2} \bigr\rangle}

\def\ie{{\em i.e.\/}\ }

\def\he#1{$^{#1}$He}

\begin{document}

\title{AlPHA MATTER REVISITED}

\author{J. W. Clark$^{1,2}$ and E. Krotscheck$^{3,4}$}

\affiliation{$^1$Centro de Investiga\c{c}\~{a}o em Matem\'{a}tica
e Aplica\c{c}\~{o}es, University of Madeira, 9020-105
Funchal, Madeira, Portugal}
\affiliation{$^2$Department of Physics, Washington University, St. Louis MO 63130, USA}
\affiliation{$^3$Department of Physics, University at Buffalo, SUNY
Buffalo NY 14260}
\affiliation{$^4$Institut f\"ur Theoretische Physik, Johannes
Kepler Universit\"at, A 4040 Linz, Austria}

\begin{abstract}
We examine in detail two alternative descriptions of a system of
$\alpha$ particles interacting via local interactions of different
character, highlighting the fact that a faithful microscopic
description of such systems demands a consistent treatment of both
short- and long-range correlations.  In preparation, we examine four
different versions of modern microscopic many-body theory and conclude
by emphasizing that these approaches, although {\it a priori} very
different, actually lead to the same equations for their efficient
application.  The only quantity that depends on the formulation of
many-body theory chosen is an {\it irreducible} interaction
correction.  In the language of Green's functions and Feynman diagrams,
it is the set of both particle-particle and particle-hole irreducible
diagrams, and in variational Jastrow-Feenberg theory it is
determined by {\it multipartite correlations} and {\it elementary
diagrams}. We apply these theoretical methods to the calculation of 
the energetics, structure, thermodynamics, and dynamics of $\alpha$
matter, as well as its condensate fraction.  In dimensionless units,
$\alpha$ matter appears to be remarkably similar to the much-studied
$^4$He quantum fluid, its low-temperature properties now basically 
solved in the Jastrow-Feenberg framework.  Accordingly, one can have 
confidence in the results of application of the same procedure to
$\alpha$ matter. Even so, closer examination reveals significant
differences between the physics of the two systems. Within an infinite
nuclear medium, alpha matter is subject to a spinoidal instability.  
Extended mixtures of nucleons and alpha particles are yet to be given 
rigorous consideration in a corresponding theoretical framework.
\end{abstract}


\maketitle

\section{Introduction}
\label{sec:introduction}

Alpha matter \cite{Tamagaki62,Nagasaki63,Harada63,ClarkWang66} was originally 
conceived as an alternative model of infinitely extended nuclear matter 
composed of intact $^4$He nuclei treated as point bosons interacting via 
a central two-body potential that fits $\alpha-\alpha$ scattering data.  
Immediately, there is the prospect of an intriguing correspondence between such
$\alpha$-particle matter and its atomic counterpart liquid \he4, for 
which the ground-state structure and low-lying excitations are now basically a 
solved quantum many-body problem.  Quite naturally, this correspondence was 
exploited in an early application of correlated-wave-function theory to $\alpha$ 
matter \cite{ClarkWang66}, as well as some of the subsequent theoretical 
investigations of the equation of state and other properties of this hypothetical 
system, now spanning nearly sixty years \cite{Khanna67}-\nocite{MuellerClark70,KhannaJairath76,ClarkJohnson78,AguileraNavarro79,JC_Kinam,MuellerLanganke94,Sedrakian96,sch,Alphanc,CarstoiuMisicu10a,CarstoiuMisicu10b,Carstoiu10,Misicu16,Satarov17,Satarov19,Zhang19,Satarov20}\cite{Satarov21}.  

Among
Refs.~\cite{Tamagaki62}\nocite{Nagasaki63,Harada63,ClarkWang66,Khanna67,MuellerClark70,KhannaJairath76,ClarkJohnson78,AguileraNavarro79,JC_Kinam,MuellerLanganke94,Sedrakian96,Alphanc,CarstoiuMisicu10a,CarstoiuMisicu10b,Carstoiu10,Misicu16,Satarov17,Satarov19,Satarov20,Zhang19}-\cite{Satarov21},r
theoretical studies of the ground state and other properties of a
system involving many interacting $\alpha$ particles divide roughly
into two categories.  In the first category, exemplified specifically
by Refs.~\cite{ClarkWang66,MuellerClark70,ClarkJohnson78} and
partially in Ref.~\cite{Alphanc}, effort is made to describe the
system at the microscopic level based on $\alpha-\alpha$ two-body
interactions that fit $\alpha-\alpha$ scattering data to a suitable
approximation.  This task is carried out by application of one or
another of the available brands of first-principles quantum many-body
theory that will be surveyed in Sec.~\ref{sec:mbt}.

The Ali-Bodmer (AB) potential \cite{AliBodmer} (1966) is the most
common choice for the basic two-body $\alpha-\alpha$ interaction in
studies of the ground-state, elementary excitations, dynamics, and
thermodynamic properties of $\alpha$ matter. With its four parameters
chosen to fit scattering data in leading states $L=0,\, 2,\, 4$ of
angular momentum up to 24 MeV, this interaction consists of an
$L$-dependent inner repulsive gaussian term and an $L$-independent
outer attractive gaussian term,
\begin{equation}
V^{(L)}(r) = V_R^{(L)}\exp{\left[-\left(\mu_R^{(L)}\right)^2 r^2\right]}   
- V_A \exp\left[-\mu_A^2r^2\right]. 
\label{eq:AliBodmer}
\end{equation}
Alternative $\alpha-\alpha$ potential models of comparable quality were 
developed earlier in the same period, also fitted to low-$L$ scattering.  
Among them, the version labeled ESH \cite{esh} features an inner hard core 
of $L$-dependent radius, plus repulsive and attractive gaussian terms of 
the same form as in the AB potential.  In the studies of $\alpha$ matter based 
on true quantum many-body theories beyond a mean-field description, the 
Ali-Bodmer interaction has generally been adopted as the standard choice for 
assessment of different microscopic many-body approaches among those methods 
reviewed in Sec.~\ref{sec:mbt}. 

The second category among theoretical approaches to prediction of the
properties of $\alpha$ matter and $\alpha$-nucleon mixtures employs
versions of the $\alpha-\alpha$ interaction {\it alternative} to the
Ali-Bodmer potential, derived by a double-folding procedure applied to
Gogny \cite{Decharge,Chappert} or Skyrme \cite{Goriely}
parametrizations of effective {\it two-nucleon} potentials
\cite{Alphanc,CarstoiuMisicu10b,Carstoiu10,Satarov17,Satarov19,Satarov20,Satarov21}.
These versions are evolved specifically from Gogny-D1 \cite{Decharge}
and Gogny-D1N \cite{Chappert} nucleon-nucleon interactions in the case
of Ref.~\cite{Alphanc}.

\begin{figure}
\centerline{\includegraphics[width=0.6\columnwidth,angle=-90]{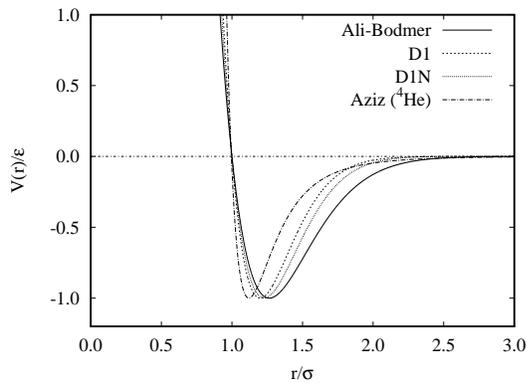}}
\caption{Aziz interaction \cite{Aziz} for \he4 (dash-dotted line) and 
Ali-Bodmer interaction \cite{AliBodmer} (solid line) for $\alpha$-matter, in  
normalized units. Also shown are curves for versions of the $\alpha-\alpha$
interaction derived from versions D1 and D1N of the Gogny variety of 
two-nucleon interactions.}
\label{fig:potplot}
\end{figure}

An explicit demonstration of the remarkable similitude of the
many-body problems of $\alpha$ matter and liquid $^4$He is provided by
Fig.~\ref{fig:potplot}.  Shown there is a comparison between the Aziz
atom-atom interaction \cite{Aziz} in liquid $^4$He and three proposed
$\alpha-\alpha$ interactions, namely the Ali-Bodmer $L=0$ interaction
and the two surrogate $\alpha-\alpha$ interactions of
Ref.~\cite{Alphanc} derived from Gogny-D1 and Gogny-D1N effective
two-nucleon interactions.

The vast difference of scales is accommodated by measuring the separation 
$r$ and potential $V(r)$ in units of the respective values of the range and 
depth parameters $\sigma$ and $\epsilon$ for the two systems. (The range 
$\sigma$ is defined as the distance below which the interaction becomes
repulsive).

For the four systems, the corresponding values of the deBoer parameter 
\cite{deBoer},
\begin{equation}
\Lambda= \left(\frac {h^2}{m\epsilon\sigma^2}\right)^{1/2}\,,
\label{eq:deBoer}
\end{equation}
which is the basis for the quantum law of corresponding 
states, are listed in Table \ref{tab:deBoer}.

\begin{table}[h!]
  \begin{center}
    \begin{tabular}{c||c|c|c|c}
      & Aziz & Ali-Bodmer &  D1 & D1N \\
      \hline
      $\sigma\ \ $ & 2.65 & 2.214 & 2.819 & 2.544 \\
      $\epsilon\ \ $ & 10.8 & 11.975 & 4.620 & 7.220 \\
      $\Lambda\ \ $ & 2.51 & 2.65 & 3.35 & 2.97\\
    \end{tabular}
    \caption{}
    \label{tab:deBoer}
  \end{center}
\end{table}

Here we must note that strictly, the deBoer parameter is defined  
specifically for 6-12 potentials.  Still, in the present case it gives 
a reasonable criterion for comparing interactions and for normalizing 
associated energetics, in support of the conclusion that $\alpha$ matter 
and liquid $^4$He are indeed rather similar.

The repulsive strength of the Ali-Bodmer interaction drops from 475
MeV for $L=0$ to 320 MeV for $L=2$ to 10 MeV for $L=4$
\cite{AliBodmer}. In our calculation of $\alpha$-matter properties, we
have used only the $L = 0$ component of this interaction, acting in
all states.  Although this simplification overestimates the binding
energy somewhat, we will find that use of the more attractive
$L$-dependent interaction would only lead to more binding and (as will
be seen) make the behavior of the model even less realistic with
respect to stability.

Our current effort belongs to the first category named above.
Sec.~\ref{sec:mbt} provides an extensive survey of state-of-the-art
quantum many-body methods, with specific attention to approaches and
aspects that are directly relevant to the microscopic physics of
$\alpha$ matter as a strongly interacting multi-boson system.  In
particular, attention is given to extended Jastrow-Feenberg theory,
parquet-diagrammatics, pair density functions, energetics,
consistency, finite-temperature behavior, and dynamics.  In
Sec.~\ref{sec:results} we present and discusses numerical results for
$\alpha$-system ground-state energetics and structure, condensate
fraction, and dynamics based on Jastrow-Feenberg and parquet theory.
Sec.~\ref{sec:conclusions} concludes with a brief summary of the
status of the alpha-matter problem.

While this paper addresses the question of the properties
and behavior of pure $\alpha$-matter, we do not attempt to solve the problem 
of how and under what conditions ordinary nuclear matter composed of a soup 
of nucleons may be subject to the formation of $\alpha$ clusters, \ie nucleon 
quartets, or the converse. Such a clustering process, considered at length 
in several review articles \cite{Yamada2011,Schuck16,PhysRevLett.80.3177}, 
is analogous to that of BCS-like pairing in neutron matter \cite{BCS50Book}. 
Just as in the BCS case, essential input its theoretical treatment is the 
interaction that causes quarteting.  Microscopic many-body theory derives 
such an interaction from an underlying microscopic interaction such as 
variants of the Reid potential \cite{Reid68,Reid93}, the Argonne 
interaction \cite{AV18}, or more modern interactions based on effective 
field theories \cite{RevModPhys.81.1773,Machleidt}. As we will see in 
the next section, this task is far less demanding than one might think. 
True, one has to deal with the complications of Fermi statistics, but 
these can be handled \cite{v3eos}-\nocite{v3twist,v3bcs,v4}\cite{v43p2}. In 
fact, for the pairing problem, serious microscopic many-body theory has
revealed effects that had been overlooked in the past, and it should
be quite worthwhile to examine analogous issues for the quarteting
problem.

\section{Generic Many-Body Theory }
\label{sec:mbt}

Microscopic many-body theory has been developed over numerous decades
since the 1950s, initually along very different lines, specifically 
quantum field theory \cite{BaymKad}, the Jastrow-Feenberg variational method
\cite{FeenbergBook}, and coupled-cluster theory \cite{KLZ}, applied
predominantly to systems of fermions.  Here we are interested in a system 
of identical bosons.  With the exception of applications within
Jastrow-Feenberg theory, Bose systems have been less well examined,
but any method developed for fermions can readily be adapted to bosons  
by taking the limit where the degree of degeneracy of single-particle 
states goes to infinity and the Fermi wave number goes to zero, while 
keeping the particle density fixed.

Certain simple physical considerations are involved in specifying
which effects a satisfactory theoretical description of an interacting
many-particle system should contain. These are:
\begin{itemize}
\item[(i)] short-range correlations to describe the influence of the
interaction on the wave function as well as saturation, and
\item[(ii)]
stability of the system under external perturbations.
\end{itemize}
Observing these simple criteria, microscopic many-body theory has been
developed along different pathways to be described briefly below. Though 
apparently very different at the outset, in the final analysis these 
approaches lead, however, to exactly the same set of equations for 
implementation.  This fact prompted the authors of Ref.~\onlinecite{parquet1} 
to conclude that ``many-body theory has been developed to a level where 
different approaches are a matter of language, but not of substance'' 
(see also Ref.~\cite{JWE93}). Accordingly, we here use the term {\em generic 
many-body theory}.

This situation is completely clear for a system of bosons. Formally, the 
case of Fermi liquids has been less extensively studied in this respect. The 
same objectives still apply, but naturally the issues become more complicated 
because of the multitude of exchange diagrams, and some additional 
approximations need to be made to establish the equivalence between the
Jastrow-Feenberg method and that based on parquet-diagram summations 
\cite{fullbcs}.

\subsection{Jastrow-Feenberg Method}
\label{ssec:JF}

Historically, the first and best-explored approach leading to what we
will call generic many-body theory is the Jastrow-Feenberg method. For bosons, 
the method starts with an {\em ansatz\/} for the wave function:
\begin{equation} 
\ket{\Psi_0} = \exp\frac{1}{2} \left[\sum_{i<j} u_2({\bf r}_i,{\bf r}_j) 
+ \sum_{i<j<k}u_3({\bf r}_{i},\rvec_j, {\bf r}_{k})+\cdot\cdot\right].
\label{eq:Jastrow}
\end{equation}
The correlation functions $u_n({\bf r}_{i_1},.., {\bf r}_{i_n})$ are
obtained by minimizing the energy expectation value:
\begin{equation}
E = \frac{\bra{\Psi_0}H\ket{\Psi_0}}
{\ovlp{\Psi_0}{\Psi_0}},\qquad
\frac{\delta E }{\delta u_n}(\rvec_1,.., \rvec_n) = 0.
\label{eq:optun}
\end{equation}
This method is, in principle, exact.  Approximations are defined by the
number of correlation functions retained and how the so-called ``elementary 
diagrams'' are treated, as explained below. The connections to the 
observable pair distribution function
\begin{eqnarray}
g(\rvec,\rvec')&=&g(|\rvec-\rvec'|)\nonumber
\\&=&\frac{1}{\rho^2}
\frac{\bra{\Psi_0}\sum_{i\ne j}\delta(\rvec_i-\rvec)\delta(\rvec_j-\rvec')
\ket{\Psi_0}}{\ovlp{\Psi_0}{\Psi_0}}
\label{eq:rho2}
\end{eqnarray}
and static structure function
\begin{equation}
 S(k) = 1 + \rho\int d^3r e^{\I{\bf k}\cdot{\bf r}}\left[g(r)-1\right]
\label{eq:sofk}
\end{equation}
are made through the hierarchy of hypernetted chain equations
\cite{Morita58,LGB}. The variations in Eqs.~\eqref{eq:optun} with respect to
the correlation functions $u_n({\bf r}_{i_1},.., {\bf r}_{i_n})$ are
then re-expressed in terms of the variations with respect to the observable
$g(\rvec-\rvec')$ and higher-order $n$-body distribution functions,
\begin{equation}
\frac{\delta E}{\delta g_n}(\rvec_1,.., \rvec_n) = 0.
\label{eq:optgn}
\end{equation}
This procedure has been described in textbooks \cite{FeenbergBook} and
pedagogical material \cite{KroTriesteBook}; we highlight here only its
most essential features. In practice, stopping at three-body correlations  
has proven to be sufficient \cite{PPA2,ChaC,EKthree}. These are dealt with
by first optimizing the triplet correlations for fixed pair
correlation functions. The result is then inserted into the energy
functional, which then only depends on $g(r)$ and $S(k)$. For further
reference, we spell out the explicit form:
\begin{equation}
E = E_{\rm R} +E_{\rm I} + E_{\rm Q},
\label{eq:etot}\end{equation}
where
\begin{eqnarray}
E_{\rm R} &=& N\frac{\rho}{2}\int d^3r \left[v(r) g(r)
+ \frac{\hbar^2}{m}\left|\nabla \sqrt{g(r)}\right|^2\right],
\label{eq:eR}\\
E_{\rm Q} &=& -\frac{N}{4} \int \frac{d^3 k}{ (2\pi)^3\rho} t(k) 
\frac{(S(k)-1)^3}{S(k)},
\label{eq:eQ}
\end{eqnarray}
in which $t(k) = \hbar^2 k^2/2m$ and $E_{\rm I}$ is the contribution 
from elementary diagrams and higher correlation functions
$u_n({\bf r}_{1},.., {\bf r}_{n})$ for $n\ge 3$.  Here, $E_{\rm I}$ is a
functional of the pair distribution function $g(r)$ that generates
the irreducible interaction through
\begin{equation}
V_{\rm I}(r)    = \frac{2}{\rho}\frac{\delta E_{\rm I}}{\delta g(r)}.
\label{eq:Virr}
\end{equation}

Manipulating the Euler equation for $g(r)$, one obtains the familiar
Bogoliubov formula
\begin{eqnarray}
S(k) &=& \frac{t(k)}{\epsilon_{\rm F}(k)}\label{eq:eFeyn}
= \left[1+\frac{2}{t(k)}{\tilde V}_{\mathrm {p-h}}(k)\right]^{-\frac{1}{2}},
\label{eq:rpa}
\end{eqnarray}
where
\begin{equation}
\epsilon_{\rm F}(k) = \sqrt{t^2(k) + 2 t(k){\tilde V}_{\mathrm {p-h}}(k)}
= \frac{t(k)}{S(k)}
\end{equation}
is the Feynman dispersion relation \cite{Feynman} and 
${\tilde V}_{\mathrm{p-h}}(k)$ is an effective local ``particle-hole''
interaction. The latter quantity is given in coordinate space by
\begin{eqnarray}
V_{\mathrm {p-h}}(r) &=& g(r)\left[v(r) + V_{\mathrm I }(r)\right] +
\frac{\hbar^2}{ m} \left|\nabla\sqrt{g(r)}\right|^2\nonumber\\
&+&\left[g(r)-1\right]w_{\mathrm I}(r),
\label{eq:Vph}
\end{eqnarray}
where $w_{\mathrm I}(r)$ is the {\em induced potential}
\begin{equation}
{\tilde w}_{\mathrm I}(k) = - \frac{t(k)}{2}\left[\frac{1}{ S^2(k)}-1\right]
-t(k)\left[S(k)-1\right].
\label{eq:wind}
\end{equation}
As usual in this field, we have defined the dimensionless Fourier transform 
by including a particle number density factor $\rho$:
\begin{equation}
{\tilde f(k)} \equiv \rho\int d^3r\, e^{\I \kvec\cdot\rvec}f(r)\label{eq:ft}.
\end{equation}
The correction $V_{\mathrm I}(r)$ in \eqref{eq:Vph} comes from the
{\em elementary diagram\/} contributions, which have to be included 
term-by-term; they change the numerical values of the results, but not 
the analytic structure of the equations. Also, three-body correlations  
lead only to a quantitative modification of that term \cite{EKthree}.

A simple rearrangement \cite{LanttoSiemens} of Eqs.~\eqref{eq:rpa} and
\eqref{eq:wind} allows us to rewrite the Euler equation in the form
\begin{equation}
\frac{\hbar^2}{ m}\nabla^2\sqrt{g(r)}
= V_{\mathrm {p-p}}(r)\sqrt{g(r)},
\label{eq:bbg}
\end{equation}
with
\begin{equation}
V_{\mathrm {p-p}}(r) \equiv v(r) + V_{\mathrm I}(r) + w_{\mathrm I}(r).
\label{eq:Vpp}
\end{equation}
Eq.~(2.15) can be recognized as a boson Bethe-Goldstone equation in
terms of an effective particle-particle interaction $V_{\mathrm p-p}(r)$.  
This observation led Buchler, Sim, and Woo \cite{Woo70} to the
conclusion that ``...it appears that the optimized Jastrow function is
capable of summing all rings and ladders, and partially all other
diagrams, to infinite order.''

It is immediately clear that the induced interaction $w_{\mathrm I}(r)$ 
has non-negligible effect in Eq.~\eqref{eq:bbg}: For $r\rightarrow\infty$, 
the {\it correct\/} pair distribution function goes \cite{FeenbergBook} 
as $g(r) \sim 1 + \hbar/(2 m c_s \rho \pi^2r^4)$, where $c_s$ is the 
hydrodynamic speed of sound. If one leaves out the correlation corrections 
$w_{\mathrm I}(r)$, the solution of Eq.~\eqref{eq:bbg} will behave as 
$g(r) \sim 1 + a/r$, where $a$ is related to the {\em S wave scattering 
length\/} of the potential. In other words, the correlation corrections 
to the particle-particle interaction must be just right to guarantee 
that $V_{\mathrm p-p}(r)$ has zero scattering length.

\subsection{Parquet Diagram Summations}
\label{ssec:parquet}

Following up on the observation of Ref.~\onlinecite{Woo70}, Jackson,
Lande, and Smith \cite{parquet1,parquet2} began with standard Green's
function perturbation theory \cite{BaymKad,FetterWalecka}. Arguing
that the self-consistent summation of ring- and ladder-diagrams was
the minimum requirement for a satisfactory microscopic description of
strongly interacting many-particle systems, they carried out the
corresponding summations and, what is most important, made the
summation practical by introducing local approximations
\cite{parquet1,parquet2}.

To summarize this incisive analysis and synthesis, the operative procedure 
amounts to the following:
\begin{itemize}
\item[(i)] Begin with a local particle-hole interaction, and sum the ring 
diagrams to obtain
\begin{equation}
\chi(k,\omega) = \frac{\chi_0(k,\omega)}{1-{\tilde V}_{\mathrm {p-h}}(k)
\chi_0(k,\omega)},
\label{eq:chi}
\end{equation}
where $\chi_0(k,\omega)$ is the density-density response function
of the non-interacting system, expressed for bosons as
\begin{equation}
\chi_0(k,\omega)=\displaystyle \frac{2 t(k)}{(\hbar\omega+\I\eta)^2-t^2(k)}.
\label{eq:Chi0Bose}
\end{equation}
The frequency integration
\begin{equation}
 S(k) = -\int_0^\infty \frac{d(\hbar\omega)}{\pi} \Im \chi(k,\omega)
\label{eq:S}
\end{equation}
then leads to the familiar Bogoliubov formula \eqref{eq:rpa}.
\item[(ii)] Define an {\em energy-dependent\/} particle-hole reducible
interaction, 
\begin{equation}
{\tilde w}_{\mathrm I}(k,\omega) = \frac{{\tilde V}_{\mathrm {p-h}}(k)}
{1-{\tilde V}_{\mathrm {p-h}}(k)\chi_0(k,\omega)}\label{eq:wkw}.
\end{equation}.
\item[(iii)] Define an {\em energy-independent\/} particle-hole reducible 
interaction 
${\tilde w}_{\mathrm I}(k)\equiv {\tilde w}_{\mathrm I}(k,\bar\omega(k))$ 
by demanding that its frequency integration gives the same observable 
$S(k)$ as the frequency integration obtained with  
${\tilde w}_{\mathrm I}(k,\omega)$; thus
\begin{eqnarray}
  &&\int_0^\infty d(\hbar\omega) \Im \left[\chi_0(k,\omega)
  {\tilde w}_{\mathrm I}(k,\omega)\chi_0(k,\omega)\right] \nonumber\\ 
  &&= \int_0^\infty (d\hbar\omega) \Im  \left[\chi_0(k,\omega)
  {\tilde w}_{\mathrm I}(k,\bar\omega(k)) \chi_0(k,\omega)\right].
\end{eqnarray}
\item[(iv)] Sum the ladder diagrams with this local interaction to arrive at
\begin{equation}
\frac{\hbar^2}{m}\nabla^2\psi(r)=\left[v(r)+w_{\mathrm I}(r)\right]\psi(r),
\end{equation}
noting that $g(r) = \left|\psi(r)\right|^2$ applies as well as 
Eq.~\eqref{eq:sofk}.
\item[(v)]
Finally, construct a local particle-hole irreducible interaction such that
the results for $S(k)$ obtained from Eq.~\eqref{eq:S} and from $g(r)$ agree.
\item[(vi)] Repeat the process to convergence.
\end{itemize}

This procedure leads to exactly the same equations \eqref{eq:rpa} and
\eqref{eq:bbg} as before, with the effective interactions \eqref{eq:Vph} and
\eqref{eq:wind}. The only difference is that the correction
term $V_{\mathrm I}(r)$ is given by the set of diagrams that are both
particle-particle and particle-hole irreducible. The equivalence
between Jastrow-Feenberg and parquet diagram summations has been
proven to the next order, with the simplest set of totally irreducible
diagrams \cite{TripletParquet} and optimized three-body Jastrow-Feenberg 
functions again leading to the same answer \cite{MixMonster}.

\subsection{Coupled-cluster method}
\label{ssec:CCM}

The coupled-cluster method (CCM), originally formulated by Coester and
K\"ummel \cite{CK} and further developed by Bishop and K\"ummel, has been 
very successful in describing electronic systems in condensed matter and 
chemical settings \cite{BiL78,Bishop,BartlettRMP,BartlettBook}.  Somewhat 
later, CCM has been applied intensively in nuclear physics, where it 
provides a plausible generalization of the Brueckner-Hartree-Fock method 
\cite{Day78,Day81,DayZab81,Hagen10,Hagen14}.

CCM is based on a ground-state ansatz the form \cite{CK}
\begin{equation}
|\Psi \rangle = e^{\cal C}|\Phi_0 \rangle,
\end{equation}  
in which $|\Phi_0 \rangle$ is a suitable reference state, notably  
for a Hartree-Fock ground state, while $\cal C$ is a cluster operator that
generates a linear combination of excited determinants from this ground
state. The exponential form ensures size extensivity of the solution.  
Relatively little has been done with the CCM for strongly interacting Bose 
systems.  In the aftermath of the work on Jastrow-Feenberg and parquet 
diagram summation and proof of their equivalence, the same issue has been 
examined within coupled-cluster theory \cite{BishopValencia}.  Importantly, 
it was established that the so-called ``Super-Sub(2)'' approximation of 
the CCM also leads to the same set of equations.

\subsection{Pair density functional theory}
\label{ssec:PairDFT}

Returning to the variational problems based on Eq.~\eqref{eq:optgn}, it is
natural to ask whether a general minimum principle exists for the pair
distribution function. In effect, we are seeking a two-body version of 
the Kohn-Hohenberg \cite{KohnHohenberg,Levy} theorem.

Following the line of arguments that led to the Kohn-Hohenberg theorem
for the one-body density, two statements can be made:
\begin{itemize}
\item[(i)] The kinetic energy $K$ depends only on $g(r)$ and not on
$v(r)$.
\item[(ii)] The total energy has a minimum equal to the ground-state
energy at the physical ground-state distribution function.  In other
words, the ground-state distribution function can be obtained through
the variational principle \eqref{eq:optgn}.
\end{itemize}

The proof of these results parallels exactly that for the original 
Kohn-Hohenberg theorem and need not be repeated here.  However, in contrast 
to the formulation of density-functional theory (DFT) where assumptions such
as the local-density approximation must be made for the energy functional, 
more is known about the properties of the pair distribution function, 
namely:
\begin{itemize}
\item[(i)] The {\it static structure function} can be derived from a
linear-response theory by the usual frequency integration \eqref{eq:S} 
from a {\it local\/} particle-hole interaction $V_{\mathrm p-h}(r)$.  Unlike 
the three cases above, Eqs.~(\ref{eq:S}) and \eqref{eq:chi} are now taken 
as the {\it definition\/} of a local particle-hole interaction consistent 
with linear-response theory.
\item[(ii)]
For small interparticle distances, the wave function should be determined 
by a two-particle Schr\"odinger equation
\begin{equation}
\hspace{0.75cm}-\frac{\hbar^2}{m}\nabla^2\Psi(r) + v(r)\Psi(r)
= \lambda\Psi(r),\quad r\rightarrow 0+,
\label{(3.1)}\end{equation}
in a very loose definition of the ``pair wave function'' $\Psi(r)$.
\end{itemize}

At short distances, the {\it pair distribution function\/} $g(r)$ is
proportional to the square of the pair wave function,
\begin{equation}
g(r) \sim \left|\Psi(r)\right|^2,\qquad r\rightarrow 0+ .
\label{(3.2)}
\end{equation}
Without loss of generality, we can therefore assume a general equation
for the pair distribution of the form \eqref{eq:bbg} for {\it all\/}
distances, where $V_{\mathrm {p-p}}(r)$ is again a {\em definition\/} of
the particle-particle interaction based on the pair distribution 
function and, accordingly, completely general.

If we now demand that the two interactions, $V_{\mathrm {p-h}}(r)$ and
$V_{\mathrm {p-p}}(r)$, yield the {\em same\/} pair distribution function
$g(r)$, we are led \cite{PairDFT} to the relationships \eqref{eq:Vph}
and \eqref{eq:Vpp}, with a yet undetermined correction term 
$V_{\mathrm I}(r)$.  The {\it new\/} aspect of the this formulation 
of the theory is the interpretation of the ``irreducible'' interaction. 
In their simplest versions, diagrammatic many-body theories start with
$V_{\mathrm I}(r)=0$ and improve upon this approximation by means of
diagram expansions \cite{ChuckReview,MixMonster,TripletParquet}. But
this is not the point from which we started here. Instead, we have
started from two rather general definitions and are, to some extent,
at liberty to choose a suitable phenomenological form of the irreducible 
interaction correction $V_{\rm I}(r)$.

\subsection{Energy Calculation}
\label{ssec:energy}

In variational theory, the equations for the pair distribution functions 
are obtained by minimizing the ground-state energy with respect to  
their variation, in particular via Eq.~\eqref{eq:optgn} for $n=2$, as 
described in Section \ref{ssec:JF}. In both the parquet-diagram and 
pair-DFT formulations, we begin with the equations of motion for the pair 
distribution function.  For this, we must assume {\em some\/} prescription 
for calculation of the irreducible interaction correction for any given 
potential, pair distribution function, and density.  Then we are able 
to calculate the pair distribution function (or the static structure 
function) for {\it any\/} potential $\lambda v(r)$ with $ 0 < \lambda < 1$.  

The Hellman-Feynman theorem \cite{hellmann1933,Feynman1939} informs us 
that the ground-state energy can be calculated by coupling-constant 
integration of the potential energy alone, simply
\begin{equation}
\frac{E}{N} = \frac{\rho}{ 2}
\int d^3r\, v(r)\int_0^1 d\lambda g_\lambda(r),
\label{eq:FH}\end{equation}
where $g_\lambda(r)$ is the pair distribution function calculated for
a potential strength $\lambda v(r)$.  The total energy then becomes
a functional of $v(r)$ and $g(r)$ of the form
\begin{equation}
\frac{E}{N} = \frac{\rho}{ 2} \int d^3r\, v(r) g(r)+ \frac{K}{N},
\label{eq:etot1}\end{equation}
where $K$ is the kinetic energy.

Now, replacing $v(r)$ by $\lambda v(r)$ in Eq.~\eqref{eq:etot} and 
differentiating with respect to $\lambda$, we have 
\begin{equation}
\frac{d}{d\lambda}\frac{E}{N}=\frac{\rho}{2}\int d^3r\,v(r)g_\lambda(r)
+ \int d^3r \left(\frac{\delta}{\delta g_\lambda}\frac{ E}{N}\right)(r)
\frac{dg_\lambda (r)}{ d\lambda}.
\label{eq:dedl}
\end{equation}
The second term in Eq.~(\ref{eq:dedl}) vanishes; hence the result
for the energy from the coupling constant integration (\ref{eq:FH}) is
the same as the energy functional. The above derivation also
shows that Eq.~(\ref{eq:FH}) is true not only for the exact ground
state, but also for any {\it approximate\/} energy functional, as long
as the pair distribution function is obtained by minimizing this
approximate energy functional.

\subsection{Consistency}
\label{ssec:consistency}

The long-wavelength limit of the structure function $S(k)$, and hence of 
${\tilde V}_{\mathrm p-h}(k)$, is determined by the the hydrodynamic speed 
of sound $c_s$, thus 
\begin{eqnarray}
S(k) &\sim& \frac{\hbar k}{2mc_s}\quad\mathrm{as}\quad 
k \rightarrow 0+,\\
{\tilde V}_{\mathrm p-h}(0+) &=& mc_s^2.
\label{eq:csfromV}
\end{eqnarray}
The hydrodynamic speed of sound can, on the other hand, be obtained from the
equation of state through
\begin{equation}
  mc_s^2 = \frac{d}{d\rho}\rho^2\frac{d}{d\rho}\frac{E}{N}
  = \rho \frac{d^2}{d\rho^2}\rho\frac{E}{N}.
\label{eq:csfromE}
\end{equation}
However, only an exact theory provides this consistency
\cite{EKVar,parquet5}.  In the present case of $\alpha$ matter we are
interested perforce in a density regime that is close to the spinodal 
density.  Accordingly, consistency between Eqs.~\eqref{eq:csfromV} and
\eqref{eq:csfromE} is imperative.  We have therefore resorted, as
described in Ref.~\onlinecite{lowdens}, to a semi-phenomenological
modification of the triplet correction to $V_{\rm I}(k)$ in the
long-wavelength regime to ensure this consistency. This does not
change the equation of state in a noticeable way.

\subsection{Finite temperatures}
\label{ssec:finiteT}

Historically, the first extension of our ``generic'' treatment of the
boson many-body problem to finite temperatures was, once again, formulated
within the Jastrow-Feenberg approach \cite{ChuckT}. In a rather
involved analysis, the entropy, and from that all other thermodynamic
quantities of interest, were calculated directly from a density-matrix 
constructed from the wave function \eqref{eq:Jastrow}.

We have indicated above how, at zero temperature, the generic many-body
equations can be derived in different ways. The same is true for the 
extension of the theory to finite temperatures: The idea is basically that 
the long-range correlations are determined by the low-lying excitations,
which are more affected by temperature.  On the other hand, short-range 
correlations are determined by the short-range interparticle interaction, 
which is temperature independent.  It is therefore legitimate to utilize 
the Bethe-Goldstone equation as a zero-temperature equation in which 
only the ``rungs,'' but not the particle-particle "ladder" propagators, 
are treated at $T>0$. We can then focus on the RPA aspect expressed in 
\eqref{eq:chi}-\eqref{eq:Chi0Bose}. The procedure required has been 
implemented within the parquet-diagram summation method 
\cite{LandeSmithT,PhysRevB.33.635}.

At finite temperature, the connection between the dynamic susceptibility 
and the dynamic structure function is
\begin{eqnarray}
S(k,\omega) &=& -\frac{1}{\pi}\frac{1}{ 1-\exp(-\hbar\omega/T)}
\Im \chi(k,\omega)\nonumber\\
&=& -\frac{1}{2\pi}\frac{ e^{\hbar\omega/2T}}{\sinh(\hbar\omega/2T)}\, 
\Im \chi(k,\omega),
\label{eq:sofkw}
\end{eqnarray}
the static structure function being just 
\begin{equation}
S(k) = \int_{-\infty}^\infty d(\hbar\omega) S(k,\omega).
\label{sofk}
\end{equation}
For bosons, the frequency integration is simple. One writes \eqref{eq:chi}
as
\begin{equation}
\chi(k,\omega)
= \frac{2t(k)}{\left(\hbar\omega+\I\eta\right)^2-\epsilon^2_{\rm F}(k)},
\end{equation}
which yields 
\begin{equation}
S(k,T) = \coth\frac{1}{2}\beta\epsilon_{\rm F}(k)S_0(k),
\label{eq:SkofT}
\end{equation}
where $S_0(k)$ is given by the expression \eqref{eq:rpa}, but where
${\tilde V}_{\rm p-h}(k)$ depends implicitly on the temperature
through the pair distribution function.

It is now straightforward to verify that the equations of motion can
be obtained from the variational principle for the free energy
\begin{equation}
F[g,n] = E_0 + \sum_\kvec\frac{\hbar^2 k^2}{2mS(k)}n_k(1+n_k) - TS,
\end{equation}
with $E_0$ given by Eq.~\eqref{eq:etot}, noting again that all
functions appearing in these expressions are temperature dependent.
Here the $n_k$ are the occupation numbers of the quasiparticle states, 
while
\begin{equation}
S = \sum_\kvec\left[(n_k+1)\ln (n_k+1) - n_k\ln n_k\right]
\label{eq:entropy}
\end{equation}
is the entropy of a Bose system with quasiparticle occupation numbers 
$n_k$.  The two independent functions $g(r)$ (or $S(k)$) and $n_k$ are 
determined by the two extremum conditions
\begin{equation}
\frac{\delta F}{ \delta g}(r) = 0\quad \hbox{and}\quad
\frac{\delta F}{\delta n}(k) = 0,
\end{equation}
which have the solutions \eqref{eq:SkofT} and
\begin{equation}
n_k = \frac{1}{\exp(\beta \epsilon_{\rm F}(k))-1}\,.
\label{eq:nofk}
\end{equation}

Considering the free energy now as a function of the scaled interaction
$\lambda v(r)$, we arrive at 
\begin{eqnarray}
\frac{d F}{ d\lambda} &=& \frac{\partial E}{ \partial\lambda}
+ \int d^3r \frac{\delta F}{\delta g}(r)\frac{d g(r)}{ d\lambda}
+ \sum_\kvec \frac{\delta F}{\delta n}(k)\frac{d n_k}{ d\lambda}
\nonumber\\
&=&
N \frac{\rho}{2} \int d^3 r v(r) g(r),
\end{eqnarray}
owing to the two optimization conditions. Hence the energy functional is 
the result of coupling-constant integration.

\subsection{Condensate fraction}
\label{ssec:nc}

Given the wave function \eqref{eq:Jastrow} and assuming that the 
correlations are known, we may now proceed to calculate the full one-body 
density matrix
\begin{eqnarray}
&&\rho_1(\rvec,\rvec') \\
&=&N \frac{\int d^3r_2\ldots d^3r_N\Psi_0(\rvec,
\rvec_2,\ldots,\rvec_N)\Psi_0(\rvec',\rvec_2,\ldots,\rvec_N)}{
\int d^3r_1\ldots d^3r_N |\Psi_0(\rvec_1,\ldots,\rvec_N)|^2}.
\label{eq:densmat}\nonumber
\end{eqnarray}
Cumulant expansions for the density matrix were derived and applied in
Refs.~\cite{RLC75,LC76,RL76} and full HNC summations carried out
in Ref.~\cite{Fantoni78}. More in line with our present approach of 
eliminating Jastrow-Feenberg type correlation functions in favor of
the pair distribution function is reformulation of the relevant
integral equations as carried out in Ref.~\cite{OldOLdSurf}. Taking
the limit of a homogeneous system, Eqs.~(5.23a)-(5.23c) of that work
become
\begin{eqnarray}
\Delta X(r) &=& \sqrt{g(r)}\exp(\Delta N(r))-\frac{1}{2}g(r)
-\frac{1}{2} -\Delta N(r), \nonumber\\
\Delta {\tilde N}(k) &=& (S(k)-1)\Delta {\tilde X}(k),
\label{eq:ncHNC}
\end{eqnarray}
with which the condensate fraction is calculated from 
\begin{eqnarray}
\ln n_c &=& 2\Delta {\tilde X}(0+)\nonumber\\
    &-&
\int\frac{d^3k}{(2\pi)^3\rho}\Delta{\tilde N}(k)
\left[\Delta{\tilde X(k)}S(k)+S(k)-1\right]\nonumber\\
&+&\frac{1}{4}\int\frac{d^3k}{(2\pi)^3\rho}\frac{(S(k)-1)^3}{S(k)}\Biggr]\,.
  \label{eq:nc}
\end{eqnarray}
These equations can be improved by adding elementary-diagram
corrections \cite{Fantoni78}, although an extension to three-body and
higher-order correlations has, to our knowledge, not been developed.

Some concerns about the validity of HNC-type expansions for the
density matrix from the standpoint of parquet-diagram summations
have been expressed in Ref.~\cite{WEJ96}; the issue has not been
investigated any further. Also, the formulation \eqref{eq:nc} leaves out
triplet and elementary diagram corrections.

\subsection{Dynamics}
\label{ssec:dynamics}

The treatment of many-body dynamics has been most extensively studied
along the lines of variational theory. One does not need to assume
explicitly a Jastrow-Feenberg wave function; it suffices to assume
that $\ket{\Psi_0}$ is the exact many-body wave function or an
approximation sufficiently close to it.

A common formulation of most treatments of the dynamics is to give the
dynamic wave function a small, time-dependent component
\begin{equation}
\left|\Psi(t)\right\rangle= e^{-\I E_0t/\hbar}
\frac{e^{\half\delta U(t)} \left|\Psi_0\right\rangle}
{\bra{\Psi_0}e^{\half\delta U^\dagger(t)}
e^{\half\delta U(t)}\ket{\Psi_0}^{1/2}}\ ,
\end{equation}
where $\ket{\Psi_0}$ is the ground state and $\delta U(t)$ is an
excitation operator, written for the case of bosons in the
form
\begin{equation}
\delta U(t) =\sum_i\delta u_1(\rvec_i;t) 
+\sum_{i<j} \delta u_2(\rvec_i,\rvec_j;t) + \ldots \,.
\label{eq:deltaU}
\end{equation}
The amplitudes $\delta u_n(\rvec_1,\ldots\rvec_;t)$ are determined
by the time-dependent generalization of the Ritz variational principle:
\begin{equation}
\frac{\delta}{\delta u_n(\rvec_1,\dots\rvec_n;t)}
\int\!\! dt\,\langle\,\Psi(t)|H-\I\hbar\partial_t|\Psi(t)\rangle = 0\,.
\label{eq:eom}
\end{equation}
This general approach and its extension to Fermi systems
\cite{2p2h,Nature_2p2h,eomIV} has been referred to as
``Dynamic Many-Body Theory (DMBT)'' and provides, to date, the most
accurate overall microscopic descrption of the dyanmics of strongly
interacting many-body systems.

For infinitesimal perturbations $\delta U(t)$ of the ground state, one
can linearize the equations of motion for $\delta u_i(\rvec_i,\dots;t)$, 
leading to the density-density response function $\chi(k,\omega)$, from 
which the dynamic structure function follows as $S(k,\omega) 
=\Im\,\chi(k,\omega)$.  Restriction of the excitation operator to one-body 
fluctuations $\delta u_1(\rvec;t)$ leads to the famous Feynman dispersion 
relation \eqref{eq:eFeyn} \cite{Feynman}, whereas a ``backflow'' choice of 
$\delta u_2(\rvec_i,\rvec_j;t)$ yields the Feynman-Cohen estimate of 
$e_0(k)$.  Unconstrained variation with respect to $\delta u_1(\rvec;t)$ and
$\delta u_2(\rvec_i,\rvec_j;t)$ gives, in a specific {\em convolution
approximation\/} for the three- and four-body vertices, the correlated-basis 
formulation of Jackson and Feenberg \cite{JaFe,JaFe2}, which is the boson 
version of what is known in nuclear physics as ``second RPA (SRPA)'' 
\cite{PhysRev.126.2231,SRPA83,SRPA87,Wambach88,PhysRevC.81.024317}. A more 
accurate evaluation of these vertices \cite{eomI} provides essentially no 
improvement.
\begin{widetext}
Permitting fluctuations $\delta u_n(\rvec_1,\ldots,\rvec_n;t)$ 
to all orders \cite{eomIII} finally leads to a response function 
of the form
\begin{equation}
\chi(k,\omega) = \frac{S(k)}{\hbar \omega - \varepsilon_{\rm F}(k) 
- \Sigma(k,\hbar\omega)} + \frac{S(k)}{-\hbar \omega 
-\varepsilon_{\rm F}(k) - \Sigma(k,-\hbar\omega)}\ ,
\label{eq:chiEOM}
\end{equation}
where the self-energy is given by an integral equation
\begin{equation}
\Sigma(k,\hbar\omega) = 
\frac{1}{2} \int \frac{d^3k_1d^3k_2}{(2\pi)^3\rho}\,
\frac{\delta({\bf k}-{\bf k}_1-{\bf k}_2)
\left|{\tilde V}_3({\bf k};{\bf k}_1,{\bf k}_2)\right|^2}
{\hbar\omega-\varepsilon_{\rm F}(k_1)
-\Sigma(k_1,\hbar\omega-\varepsilon_{\rm F}(k_2))
-\varepsilon_{\rm F}(k_2)-\Sigma(k_2,\hbar\omega
-\varepsilon_{\rm F}(k_1))},
\label{eq:sigma}
\end{equation}
in which ${\tilde V}_3(\kvec;\pvec,\qvec)$ is the three-phonon vertex
\begin{equation}
{\tilde V}_3({\bf k};{\bf k}_1,{\bf k}_2)
 =\frac{\hbar^2}{2m}\sqrt{\frac{S(k_1)S(k_2)}{S(k)}}
\left[{\bf k}\cdot {\bf k}_1{\tilde X}(k_1)+{\bf k}\cdot {\bf k_2}
{\tilde X}(k_2) - k^2{\tilde X}_3({\bf k},{\bf k}_1,{\bf k}_2)\right]\,,
\label{eq:V3}
\end{equation}
\end{widetext}
where ${\tilde X}(k)=1-1/S(k)$ and ${\tilde X}_3({\bf k},{\bf k}_1,
{\bf k}_2)$ is the fully irreducible three-phonon coupling
matrix element. In the simplest approximation, ${\tilde X}_3({\bf k},
{\bf k}_1,{\bf k}_2)$ is replaced by the three-body correlation
${\tilde u}_3(\kvec,\kvec_1, \kvec_2)$.  This approximation ensures
that long-wavelength properties of the excitation spectrum are
preserved \cite{Chuckphonon}. The improved calculations mentioned
above \cite{eomI} sum a 3-point integral equation to guarantee satisfaction
of exact properties of ${\tilde X}_3({\bf k},{\bf k}_1,{\bf k}_2)$ as $k
\rightarrow 0+$ {\em and\/} of the Fourier transform 
$X_3({\bf r}_1,{\bf r}_2,{\bf r}_3)$ for $|{\bf r}_1-{\bf r}_2| \rightarrow 0$
and $|{\bf r}_1-{\bf r}_3| \rightarrow 0$ \cite{eomI}. 

We include these corrections routinely; they have a small
but visible effect only for wave vectors between the maxon and the
roton. The Correlated Basis Functions Brillouin-Wigner (CBF-BW) approximation
\cite{JacksonSkw} is obtained by omitting the self-energy corrections
in the energy denominator of Eq.~(\ref{eq:sigma}).

\subsection{Summary of theory}
\label{ssec:TheorySummary}

We have outlined above four different ways to arrive at exactly the
same set of basic equations for a strongly interacting many-body
system.  At this juncture we need to stress the simplicity of the
method. All it takes is the iterative solution of
Eqs.~\eqref{eq:rpa}-\eqref{eq:wind} or, alternatively,
Eqs.~\eqref{eq:bbg}, \eqref{eq:Vpp}, and \eqref{eq:wind}.  In
particular, both equations \eqref{eq:rpa} and \eqref{eq:bbg} have been
at the center of many-body theory for decades. One simply asks that
the induced interaction is determined such that their solution are the
same. The only quantity that requires either diagrammatic or
phenomenological input is the irreducible interaction correction
$V_{\rm I}(r)$. Even the simplest choice, $V_{\rm I}(r) = 0$, recovers
about 80 percent of the binding energy in liquid \he4 and, as we shall
see, around 90 percent of the binding energy in $\alpha$ matter.

\section{Results}
\label{sec:results}
\subsection{Energetics and Structure}
\label{ssec:structure}

\begin{figure}
\centerline{\includegraphics[width=0.6\columnwidth,angle=-90]
{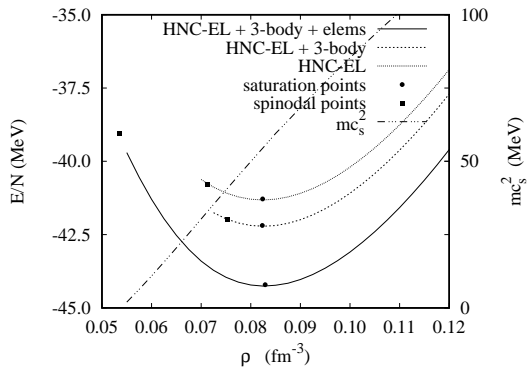}}
\caption{Comparison between results for the 
equation of state of alpha matter for the Ali-Bodmer interaction 
within the HNC framework in HNC-EL//0 approximation as well as when 
triplets and elementary diagrams are included (left scale).  
Also shown is the long-wavelength limit, with $mc_s^2 = 
{\tilde V}_{\mathrm p-h}(0+)$ (right scale).}
\label{fig:ABeos}
\end{figure}

\begin{figure}
\centerline{\includegraphics[width=0.6\columnwidth,angle=-90]
{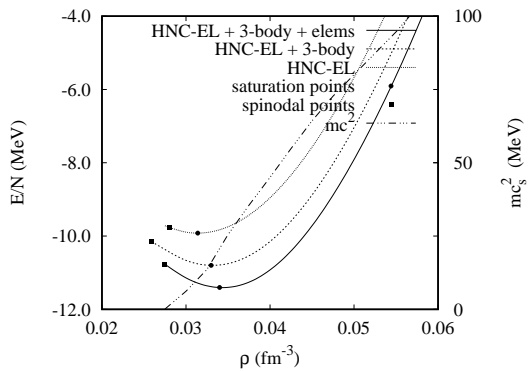}}
\caption{Same as Fig.~\ref{fig:ABeos} but for the D1N interaction.}
\label{fig:Gognyeos}
\end{figure}

The saturation density (per nucleon) of isospin-symmetric nuclear
matter is $0.160\,$fm$^{-3}$ \cite{PhysRevC.58.1804}, which
corresponds to an $\alpha$-particle density of $0.04\,$fm$^{-3}$. As
will be seen, the latter value is below both the spinodal density
$0.053\,\mathrm{fm}^{-3}$ and the saturation density of
$0.082\,\mathrm{fm}^{-3}$ of $\alpha$-matter as predicted for the
Ali-Bodmer potential (cf.~Fig.~\ref{fig:ABeos}).  We have also
calculated the equation of state for the $\alpha-\alpha$ interactions
generated from the D1 and D1N versions Gogny interactions by the
double-folding procedure applied in Ref.~\onlinecite{Alphanc}.  The
version D1 did not lead to binding; accordingly its results are not
shown. Version D1N led to a binding energy per particle of $-11.4$ MeV
at a saturation density of $0.034\,\mathrm{fm}^{-3}$, as seen in
Fig.~\ref{fig:Gognyeos}.  While the saturation density is more
reasonable, the binding energy is far too small to provide a faithful
model for generic nuclear matter.

Information on the reliability of our calculations can be obtained by
considering the convergence of the HNC-EL calculations for $\alpha$
matter. Even the simplest approximation, $V_{\rm I}(r) = 0$, recovers
about 90\,\% of the binding energy at saturation density. Our
results in the HNC-EL//0 approximation agree generally quite well with
the results of Ref.~\onlinecite{JC_Kinam}.  Presumably due to limited 
computational resources then available, that early work only missed
the fact that HNC-EL//0 has no solution below the spinodal density.

We have already stressed the similarity of alpha matter to $^4$He. In fact, 
one sees here that the convergence of energy calculations is much better 
than in $^4$He, where the HNC-EL//0 approximation recovers only about 75\,\% 
of the binding energy \cite{EKthree}.

\begin{figure}
\centerline{\includegraphics[width=0.6\columnwidth,angle=-90]
{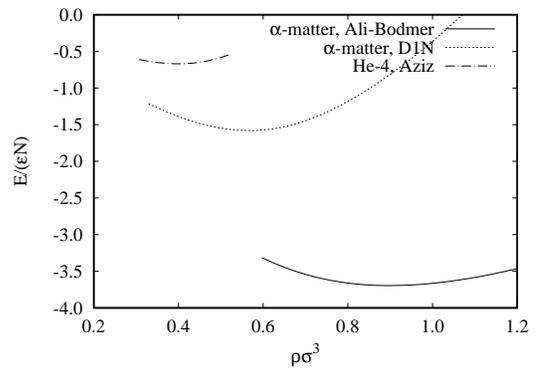}}

\caption{Comparison of Ali-Bodmer (solid line) and D1N (dashed line) 
equations of state for alpha matter with that of liquid $^4$He for the 
Aziz interaction (dash-dotted line), in dimensionless units, energy 
in units of potential depth $\epsilon$, and lengths in units of core 
size $\sigma$.}
\label{fig:eoscompare}
\end{figure}

A rather different picture emerges when comparing equations of state
expressed in dimensionless units, as suggested by deBoer
scaling. According to the values of the respective deBoer parameters
$\Lambda$ (Eq. \eqref{eq:deBoer} and Table \ref{tab:deBoer}, the three
equations of state should be quite similar when expressed in these
units.  Quite evidently, Fig.~\ref{fig:eoscompare} shows that they are
not.  While the scaled orders of magnitude are still comparable,
$\alpha$ matter is much more strongly bound and has a saturation
density three times higher.  The source of this disparity is the much
broader attractive region of the $\alpha-\alpha$ interactions as
compared to the Aziz He-He potential.

\begin{figure}
\centerline{\includegraphics[width=0.6\columnwidth,angle=-90]
{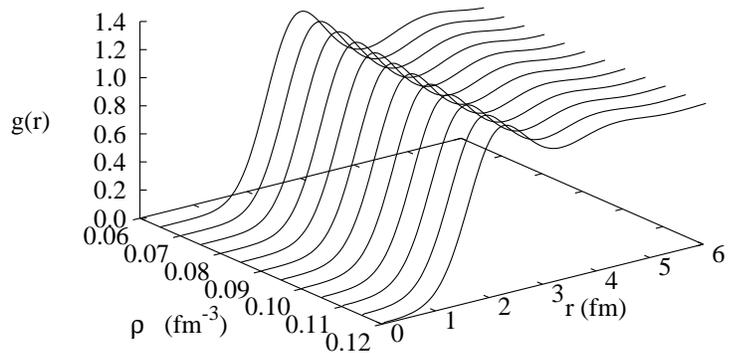}}
\caption{Pair distribution function $g(r)$ as a function of 
density $\rho$ for the Ali-Bodmer interaction.} 
\label{fig:ABgofr}
\end{figure}

\begin{figure}
\centerline{\includegraphics[width=0.6\columnwidth,angle=-90]
{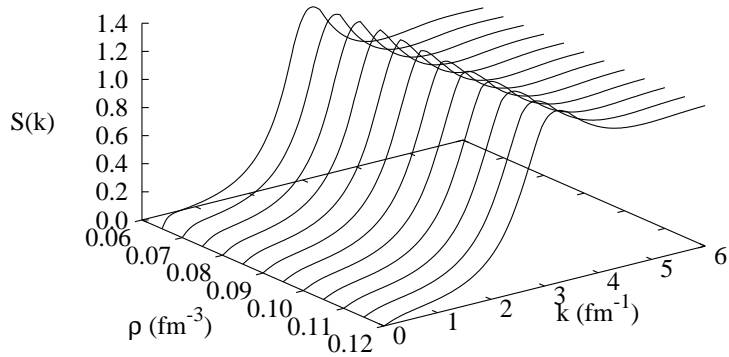}}
\caption{Static structure function $S(k)$ as a function of density
$\rho$ for the Ali-Bodmer interaction.} 
\label{fig:ABSofq}
\end{figure}

For completeness, Figs.~\ref{fig:ABgofr} and \ref{fig:ABSofq} show the
pair distribution function and the static structure function obtained
for alpha matter as a function of density.  Qualitatively, the results
are not very different from those for liquid $^4$He. However, a remarkable
feature of alpha matter is that, unlike the situation in $^4$He, the
``nearest neighbor peak'' in both $g(r)$ and $S(k)$ seems to {\em
decrease\/} as a function of density, indicating that, remarkably, 
the system becomes {\it more strongly correlated\/} when the density 
is lowered.

We conclude this section with a comment on the importance of 
``beyond parquet'' corrections that are represented by the interaction
correction $V_{\rm I}(r)$ and the energy correction $E_{\rm I}$.
We have seen above that already the simplest approximation,
$E_{\rm I}=0$, gives a very good prediction for the ground-state
energy. Caution should be exercised, however, when generalizing that
statement to other quantities. The error in the energy, or, more generally
the error in all quantities that follow from a variational principle,
is {\em quadratic\/} in the deviation of the approximate wave function
from the exact one. This applies to all quantities considered here
{\em except,\/} as we shall see, the condensate fraction, which depends
sensitively  on the pair distribution function.

\begin{figure}
\centerline{\includegraphics[width=0.6\columnwidth,angle=-90]
{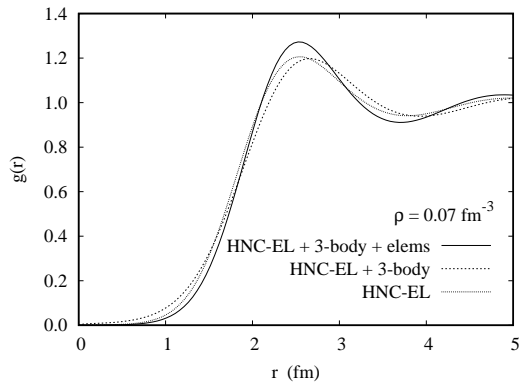}}
\caption{Pair distribution function $g(r)$ at the density $\rho =
0.07\,$fm$^{-3}$, for the Ali-Bodmer interaction. The solid line shows 
the result including 3-body and elementary-diagram corrections; the 
dashed line, the result including only 3-body correlations; and 
the dotted line, the HNC-EL//0 approximation.} 
\label{fig:ABgofr07}
\end{figure}

\begin{figure}
\centerline{\includegraphics[width=0.6\columnwidth,angle=-90]
{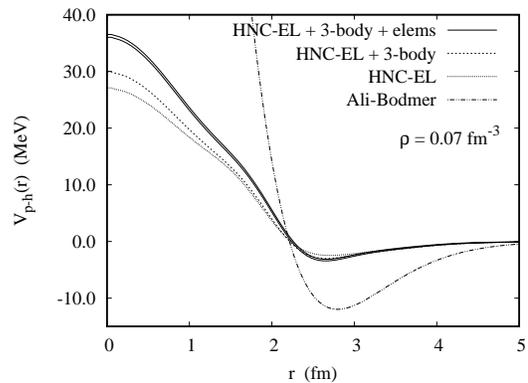}}
\caption{Same as in Fig.~\ref{fig:ABgofr07}, but for the particle-hole
interaction $V_{\rm p-h}(r)$. The two solid lines show $V_{\rm p-h}(r)$ 
when the semi-phenomenological modifications discussed in Section 
\ref{ssec:consistency} are respectively included or omitted.  Also shown 
is the bare Ali-Bodmer potential (dash-dotted line).
\label{fig:ABvph07}}
\end{figure}

Figs.~\ref{fig:ABgofr07} and \ref{fig:ABvph07} show three different 
approximations to the pair distribution function $g(r)$ for the 
Ali-Bodmer potential. It was necessary to go to rather high density, 
$\rho = 0.07\,$fm$^{-3}$, because the HNC-EL//0 approximation has no 
solution at lower densities.

An interesting feature is seen in the particle-hole interaction, which 
again sets $\alpha$-matter apart from $^4$He. Phenomenologically it is 
argued \cite{Aldrich} that in liquid $^4$He the quantity $V_{\rm p-h}(r)$
(called the ``pseudpotential'' by Aldrich and Pines) should display:
\begin{itemize}
\item[(i)] an enhancement of the short-distance repulsion due
to the cost in kinetic energy for bending the wave function
to zero at small interpartcle distances, and
\item[(ii)] an enhanced attraction at the potential minimum due 
to the presence of neighboring attractive particles.
\end{itemize}
In contrast to $^4$He where these effects are faithfully reproduced,
they are not seen in our results for $\alpha$-matter.

\subsection{Finite Temperature}

\begin{figure}[h]
\centerline{\includegraphics[width=0.6\columnwidth,angle=-90]
{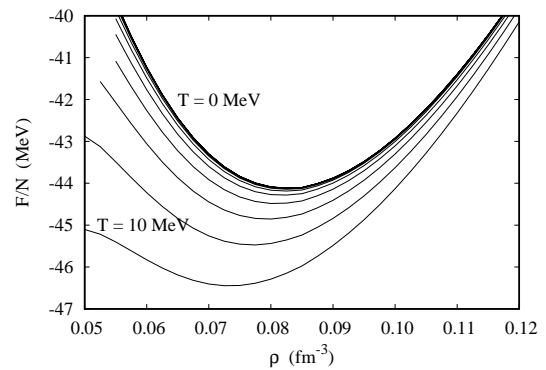}}
\caption{Free energy $F/N$ per particle for the Ali-Bodmer interaction 
in the temperature range $k_{\rm B}T  = 0, 1, \ldots 10~$MeV.
\label{fig:ABfree}}
\end{figure}

Results for our calculations at finite temperature are shown in
Figs.~\ref{fig:ABfree} and \ref{fig:GNfree}. We have limited these
calculations to the regime $ 0 \le k_{\rm B} T \le 10$~MeV, because
the theory formulated in section \ref{ssec:finiteT} assumes a Feynman 
spectrum that has a roton minimum between 30 and 40 MeV, whereas the 
best prediction for the collective excitations suggests a roton minimum
of 20 to 30 MeV. (See Subsection \ref{ssec:V.A}.)  Experience 
from $^4$He indicates that rotons already contribute visibly to the 
thermodynamics of the system at about a tenth of the roton energy 
\cite{Landau5}; for a quantitative analysis, see Fig.~50 of 
Ref.~\onlinecite{He4Dispersion}.

\begin{figure}[h]
\end{figure}\begin{figure}[h]
\centerline{\includegraphics[width=0.6\columnwidth,angle=-90]
{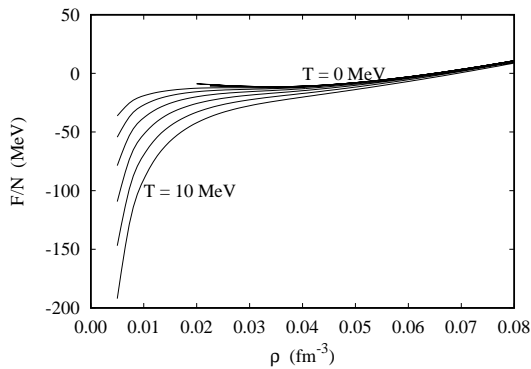}}
\caption{Same as Fig.~\ref{fig:ABfree}, but for the D1N interaction.
Note that the fact that the higher-temperature curves of $F/N$ are concave
does not contradict the stability condition $mc_s^2 > 0$, since $\rho F/N$ 
must be convex (see Eq.~\eqref{eq:csfromE}.
\label{fig:GNfree}}
\end{figure}

An interesting feature that sets $\alpha$-matter apart from the
otherwise rather similar $^4$He fluid is that, within the temperature 
regime studied, we did not observe a significant change of the spinodal
density, whereas in $^4$He \cite{RistigT} this feature is already observed 
at 4~K. There is a slight bending of the spinodal line toward lower
densities above ~10\,MeV, but we did not pursue that behavior any
further. Improvements of the finite-termperature theory \cite{CKSC}
lead to the replacement of the Feynman spectrum by the one predicted
by (CBF-BW) approximation \cite{JacksonSkw}.

\subsection{Condensate fraction}

\begin{figure}[h]
\centerline{\includegraphics[width=0.6\columnwidth,angle=-90]{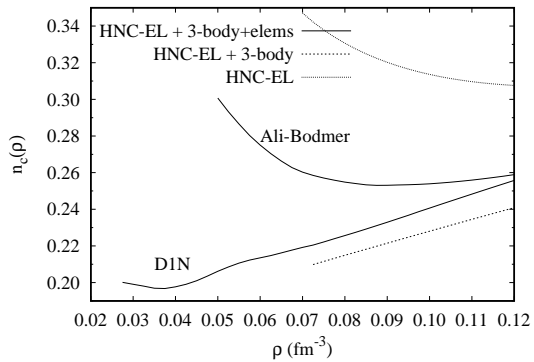}}
\caption{Condensate fraction of $\alpha$-matter as a function of 
density. The reference to ``3-body and elems'' refers to the input function 
$g(r)$, but no three-body correlations or elementary diagrams were retained 
in the explicit expression, \ie Eqs.~\eqref{eq:ncHNC} and \eqref{eq:nc}
have been used for all calculations. The labeled solid lines depict the
most complete calculations for the two interactions considered in this
work; simpler versions for the D1N interaction are not shown.
\label{fig:ABnc}}
\end{figure}

Results for the condensate fraction $n_c$ of Eq.~\eqref{eq:nc} are
shown in Fig.~\ref{fig:ABnc}. One can improve upon this calculation by
including irreducible ``elementary-'' or ``triplet correlation-''
diagram corrections in Eq.~\ref{eq:nc}.  We have deliberately not 
included these corrections, in order to demonstrate the sensitive 
dependence of the results on the input pair correlation function.

The results for the Ali-Bodmer interaction are in reasonable
agreement with those of Ref.~\onlinecite{Alphanc}, although the latter
work uses rather simple cluster expansions and correlation functions.
We also show results for the cases where elementary diagrams and/or
three-body correlations are omitted. Obviously, the results are quite
different and underscore our statement above on the sensitive dependence
of the condensate fraction on the input data. For example, one may compare 
the results in Fig.~\ref{fig:ABgofr07} with the values of $n_c$ at the 
same density.  Our results do not agree well with those of 
Ref.~\onlinecite{Alphanc} for the D1N interaction. Evidently, the reason 
for this disagreement also lies in the fact that the condensate 
fraction depends sensitively on the pair correlation function.

The expected accuracy of our results is therefore similar to that of
the energy calculation. One can also judge their accuracy by comparing
HNC-type calculations \cite{Indian85} for \he4 with corresponding
Monte Carlo calculations \cite{JordiQFSbook,MoroniCondensate,Jordinc},
although the correlation functions of Ref.~\onlinecite{Indian85}
were also non-optimized.

\subsection{Dynamics}
\label{ssec:V.A}

\begin{figure}[h]
\centerline{\includegraphics[width=0.6\columnwidth,angle=-90]{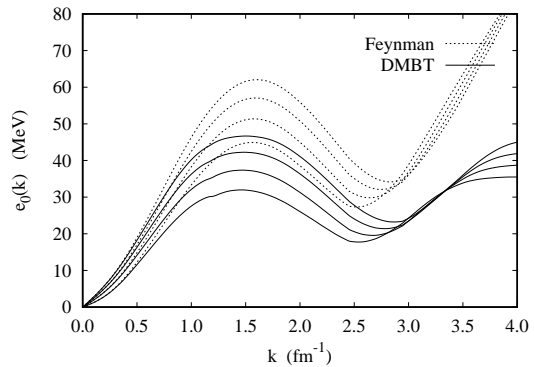}}
\caption{Zero-sound dispersion relation $e_0(k)$ for the Ali-Bodmer
  interaction in Feynman approximation (long-dashed lines), and DMBT
  \cite{eomIII} (solid lines), for the densities $\rho = 0.06, 0.07,
  0.08, 0.09\,\mathrm{fm}^{-3}$.  The highest density corresponds to
  the highest roton minimum.}\label{fig:ezsplot}
\end{figure}

Implementation of the method outlined only briefly in section
\ref{ssec:dynamics} has led to an unprecedented agreement between
theoretical predictions \cite{eomIII} and experimental results
\cite{skw4lett,He4Dispersion} in \he4. A small quantitative
improvement can be obtained by including 4-body CBF corrections
\cite{LeeLee}; these have not been included in the present 
application. The quantity of most immediate interest is the phonon 
dispersion relation $e_0(k)$, which is given by the pole of the 
density-density response function \eqref{eq:chiEOM}. Our results for 
a sequence of densities around the equilibrium density for the 
Ali-Bodmer interaction are shown in Fig.~\ref{fig:ezsplot}. A 
feature that immediatly distingushes our results for $\alpha$ matter 
from those for \he4 is that both the energy of the roton minimum and 
that of the maxon are found to {\em increase\/} rather than decrease 
with density. This behavior consistent with the fact, pointed out above,
that the behavior of the nearest-neighbor peak in $g(r)$ also
indicates that the lower-density system is more strongly correlated.

\begin{figure}[h]
\centerline{\includegraphics[width=0.6\columnwidth,angle=-90]{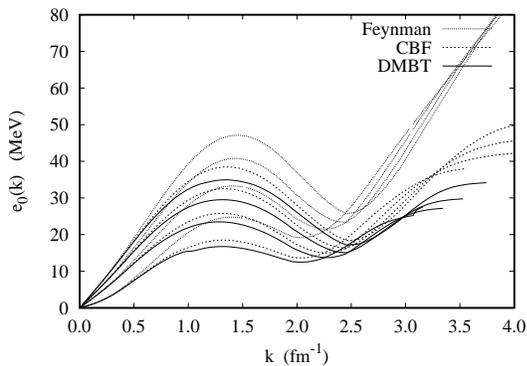}}
\caption{Same as Fig.~\ref{fig:ezsplot}, but for the D1N interaction and  
densities $\rho = 0.03, 0.04, 0.05, 0.06\,\mathrm{fm}^{-3}$. The highest 
density corresponds to the highest roton minimum.}\label{fig:D1Nezsplot}
\end{figure}

Our results for the D1N interaction shown in Fig.~\ref{fig:D1Nezsplot}
basically support these results. The ``roton minimum'' appears at a
somewhat larger wave length, which is expected because the equilibrium 
density is substantially lower.

\begin{figure*}[h]
\centerline{
\includegraphics[width=0.27\textwidth,angle=-90]{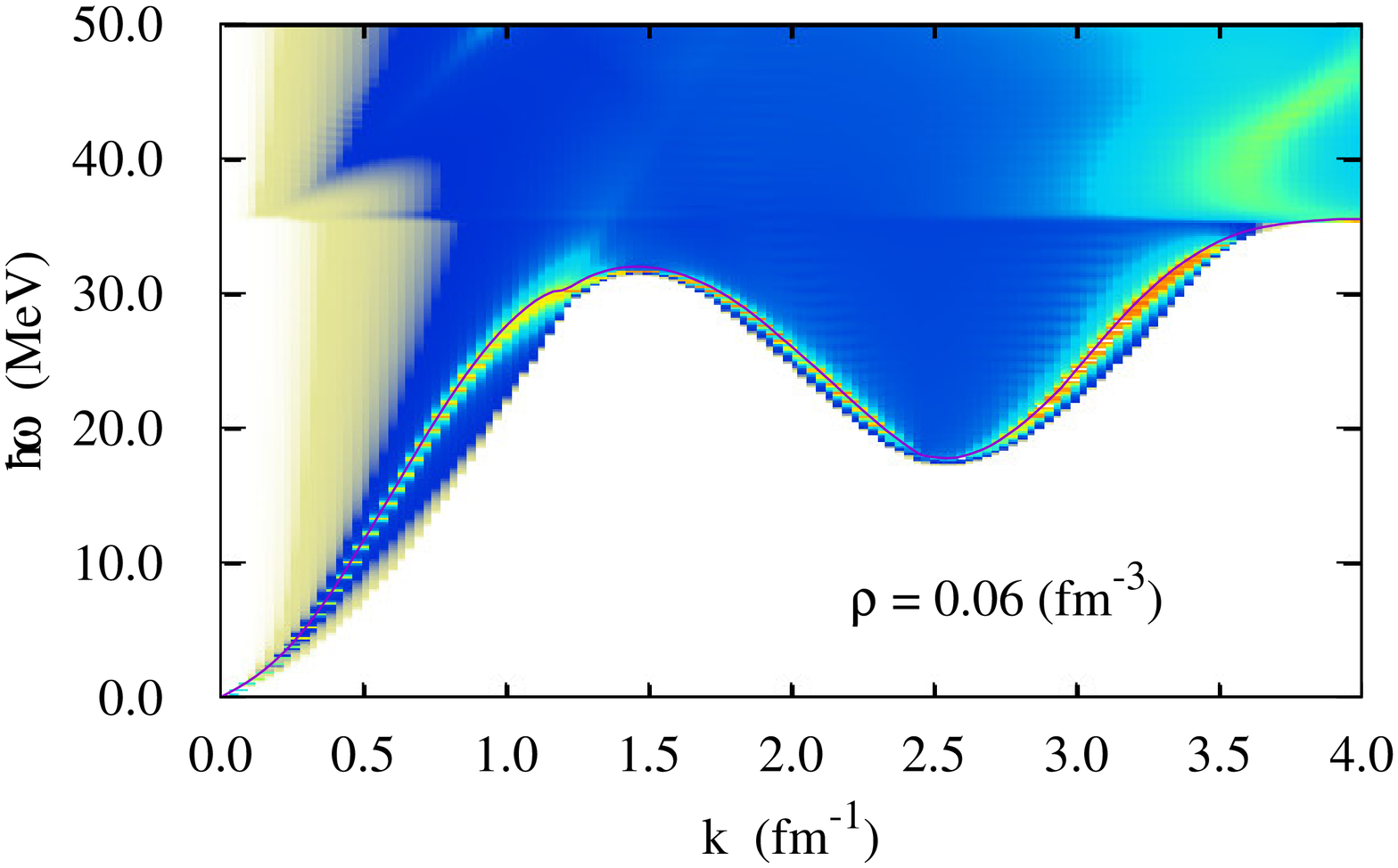}
\includegraphics[width=0.27\textwidth,angle=-90]{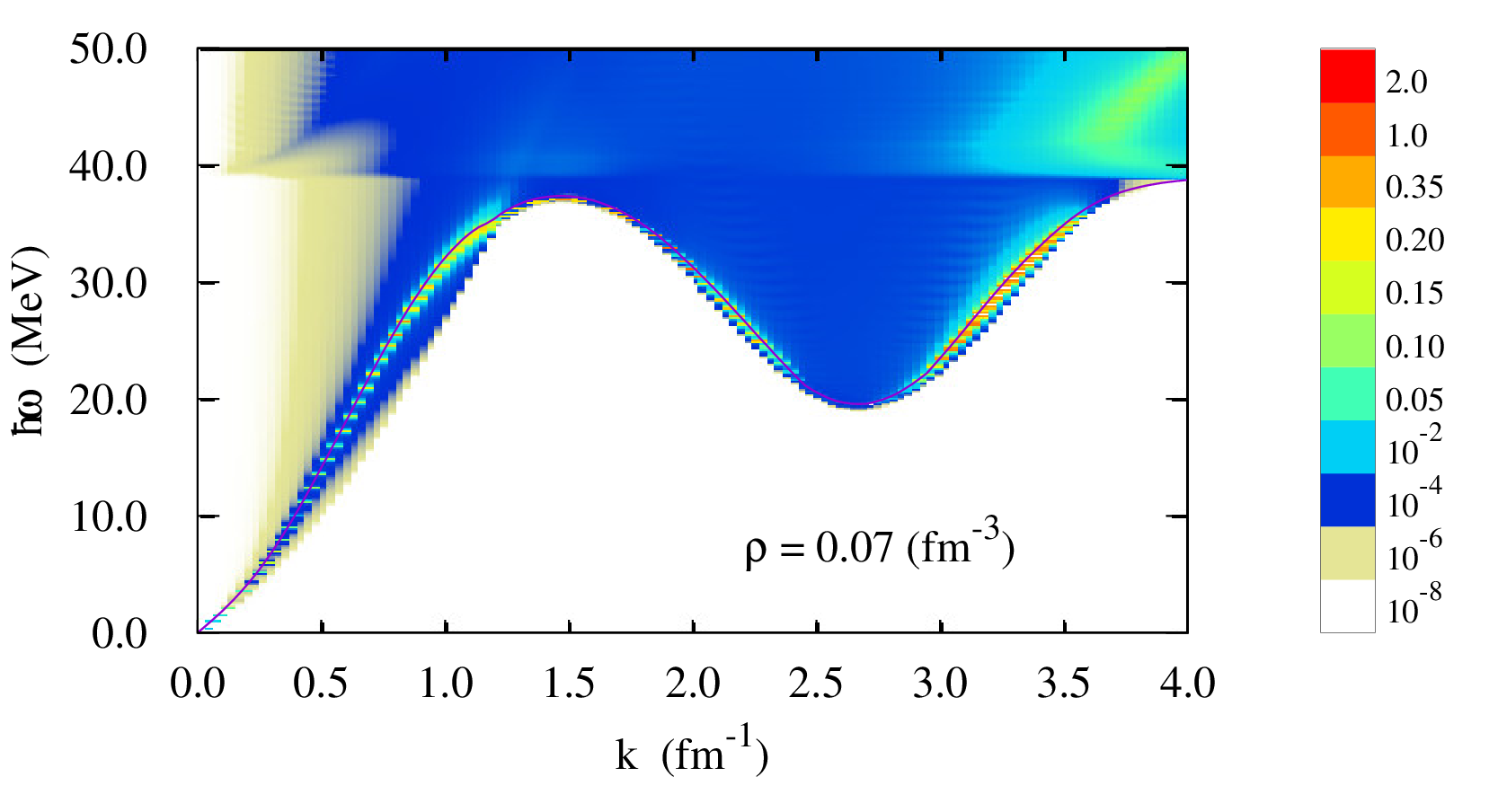}}
\centerline{
\includegraphics[width=0.27\textwidth,angle=-90]{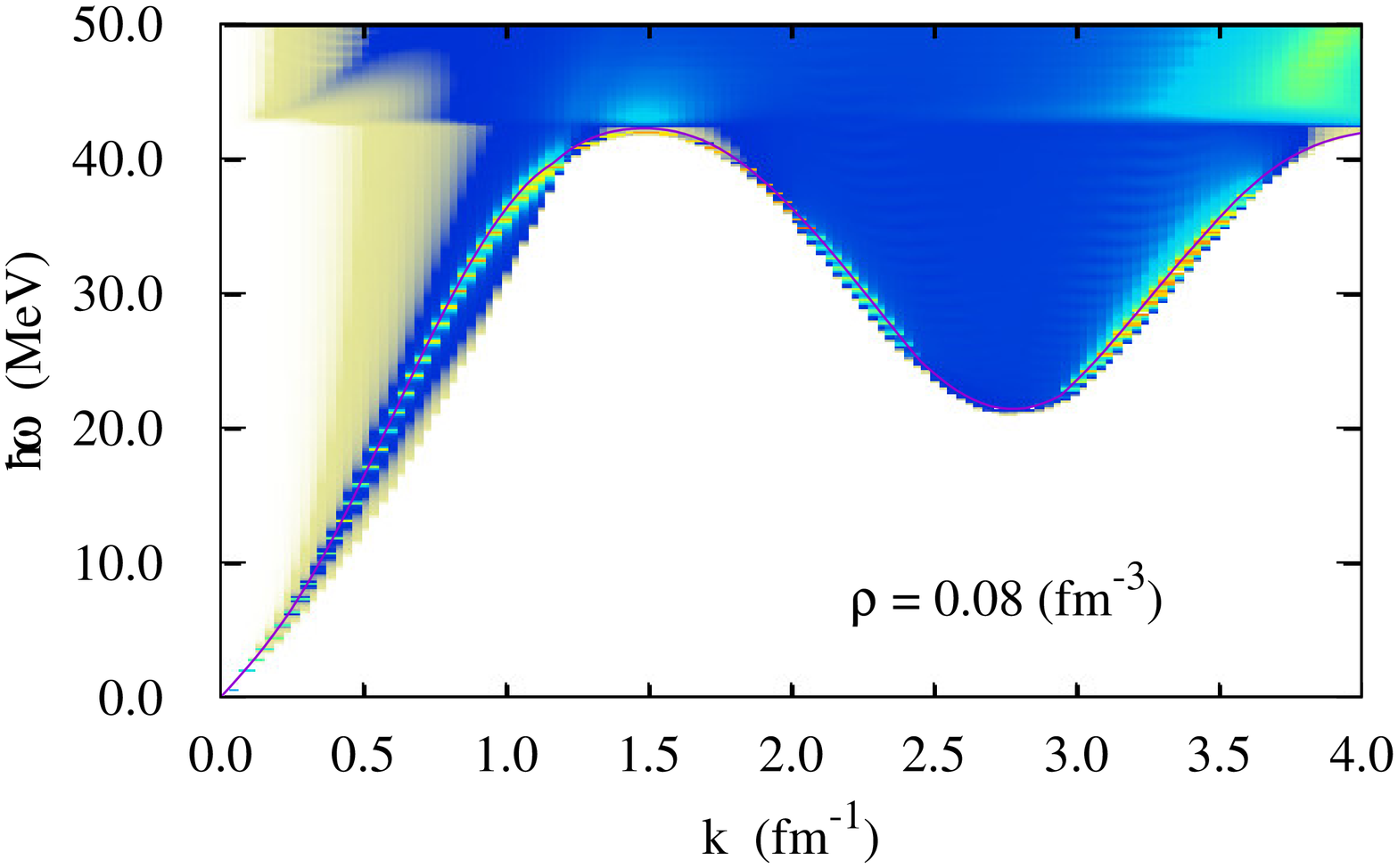}
\includegraphics[width=0.27\textwidth,angle=-90]{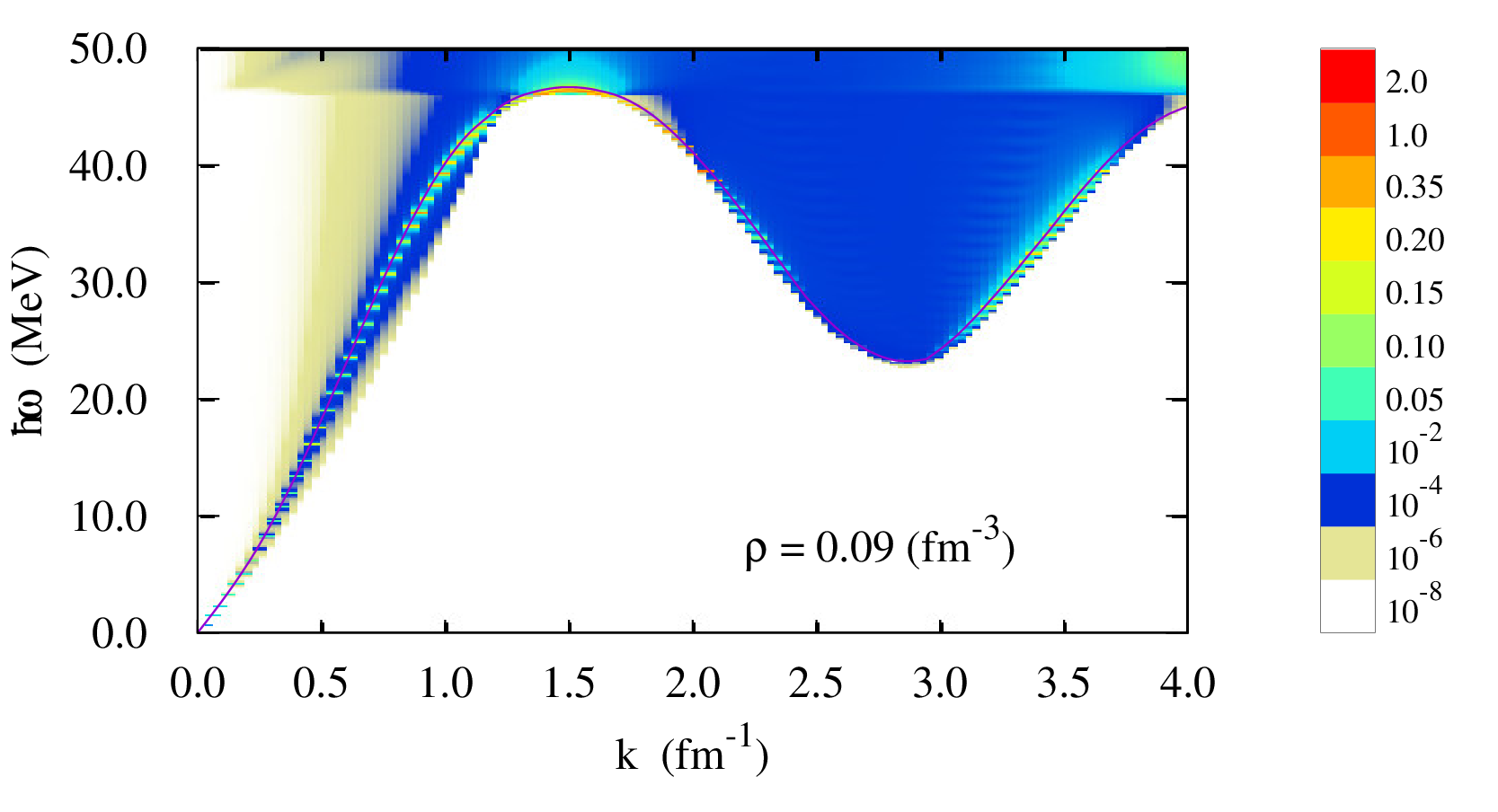}
}
\caption{(color online) Maps of $S(k,\omega)$ for the Ali-Bodmer
potential for a sequence of densities
$\rho=0.06\,\mathrm{fm}^{-3}$,\dots,\,$\rho=0.09\,\mathrm{fm}^{-3}$.
The solid red line is the phonon dispersion relation, also shown in
Fig.~\ref{fig:ezsplot}.\label{fig:skwplotcolors}}
\end{figure*}

\begin{figure*}[h]
\centerline{
\includegraphics[width=0.27\textwidth,angle=-90]{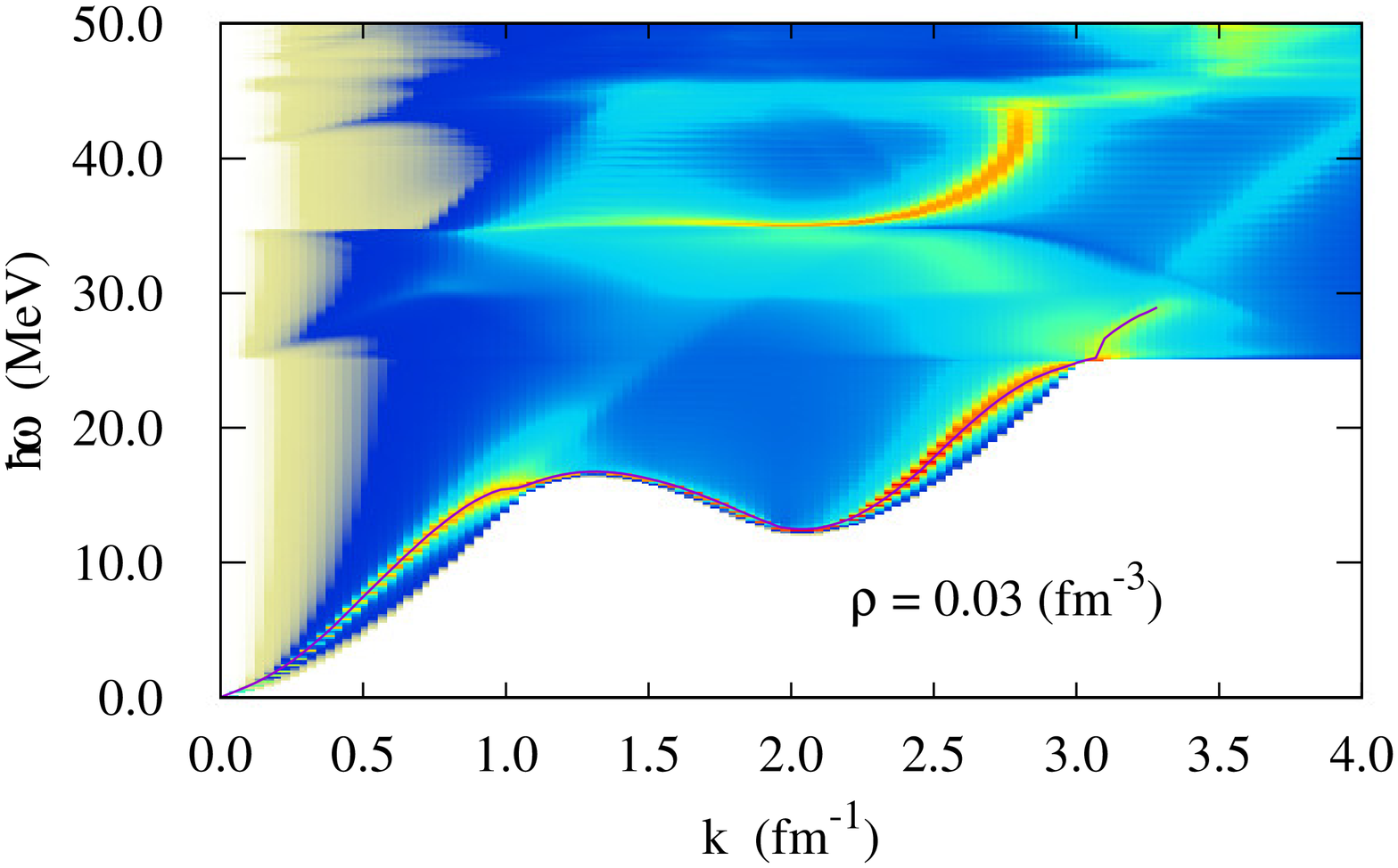}
\includegraphics[width=0.27\textwidth,angle=-90]{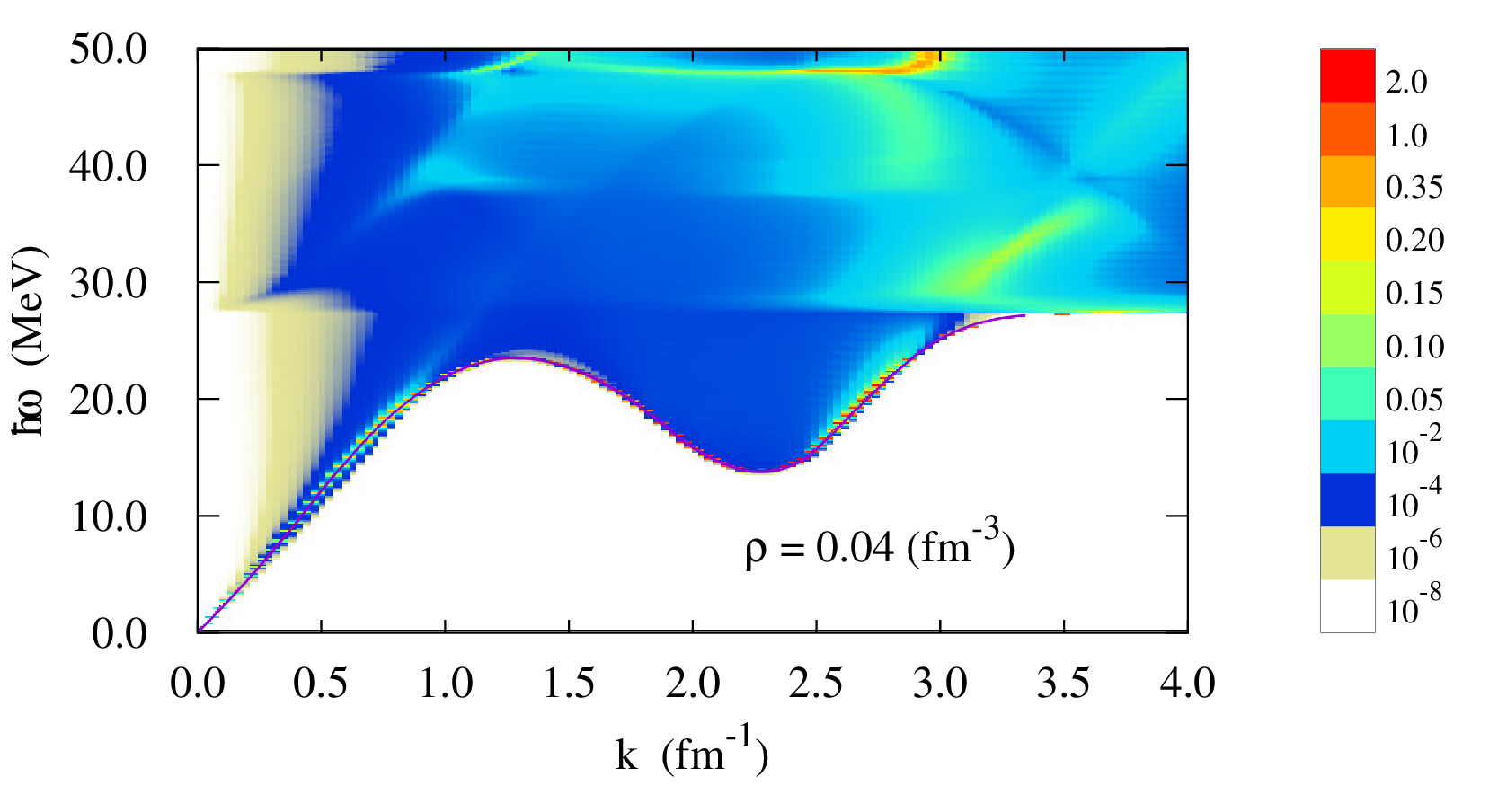} }
\centerline{
\includegraphics[width=0.27\textwidth,angle=-90]{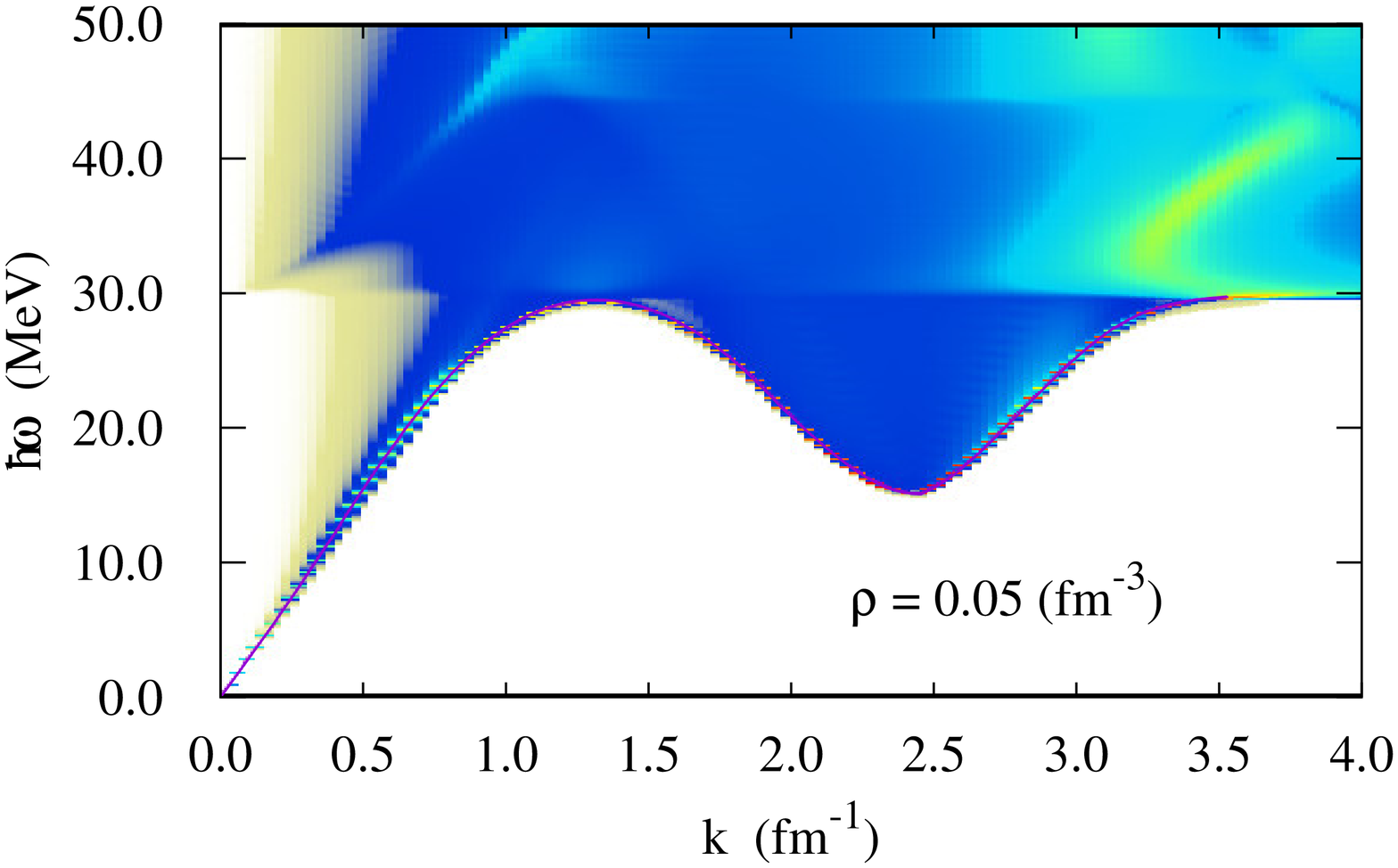}
\includegraphics[width=0.27\textwidth,angle=-90]{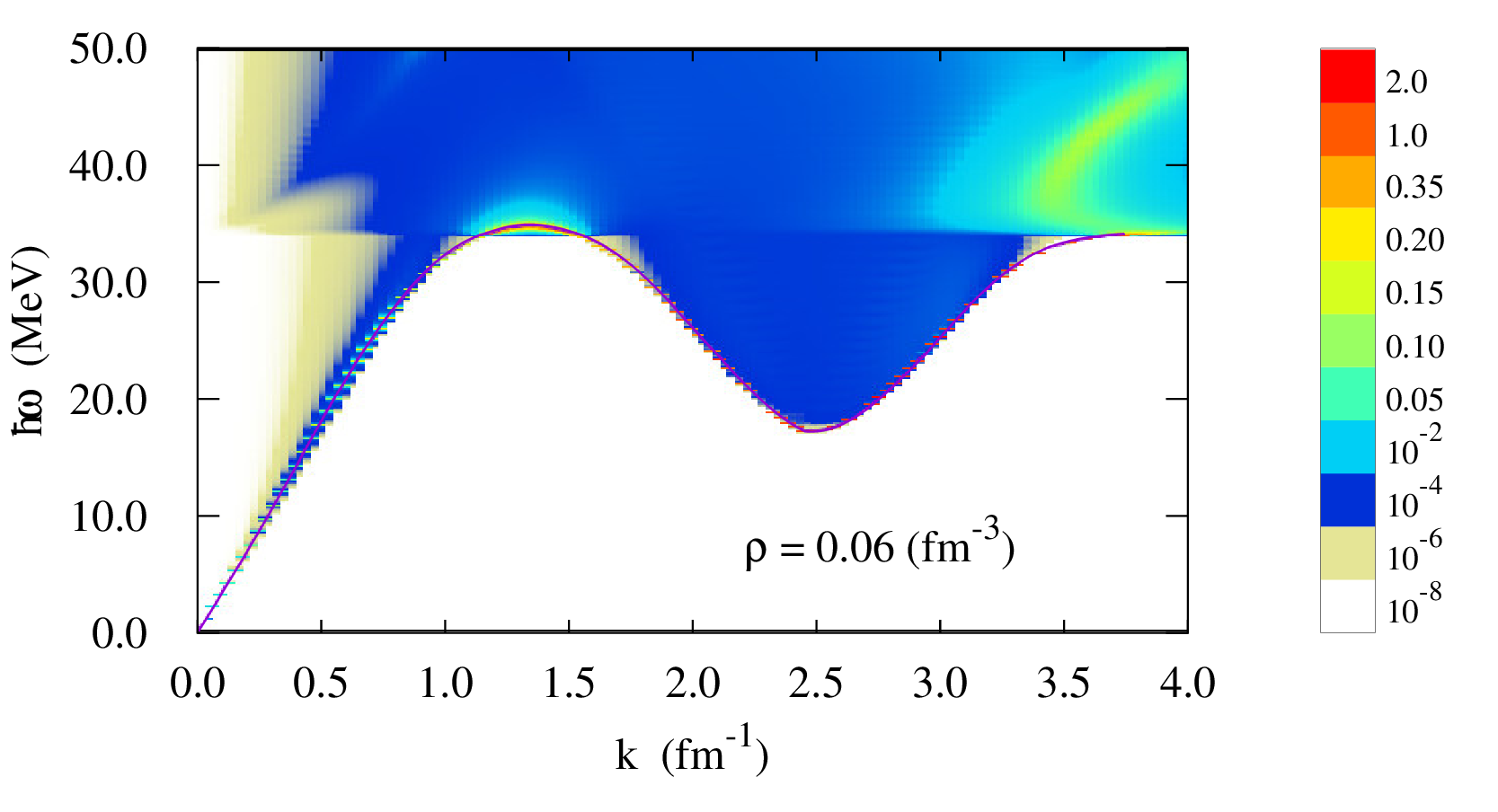} }
\caption{(color online) Same as Fig. \ref{fig:skwplotcolors} for the
D1N interaction and a sequence of
densities $\rho=0.03\,\mathrm{fm}^{-3}$,\dots,\,$\rho=0.06\,\mathrm{fm}^{-3}$.
\label{fig:D1Nskwplotcolors}}
\end{figure*}

The phonon-roton dispersion relation is only a part of the story: From
the fact that we are close to the spinodal point we can conclude that
the phonon dispersion relation is anomalous, which has the consequence
that the phonon has a finite width. Moreover, the existence of a
typical phonon-maxon-roton structure has the consequence that for
larger momentum transfers, there should be a ``Pitaevskii Plateau''
\cite{Pitaevskii2Roton}.  All of these features are seen in the
contour plots shown in Figs.~\ref{fig:skwplotcolors} and
\ref{fig:D1Nskwplotcolors}. In fact, some of the features exhibited,
such as the extension of the plateau to long wavelengths and the
extension of the $R_+$ roton to higher energies as well as its
Cherenkov damping, are seen more clearly than in \he4.

Fig.~\ref{fig:D1Nskwplotcolors} provides the same information for the D1N
interaction. At the lowest densities, $S(k,\omega)$ exhibits a remarkably
rich structure at high energies, which can be attributed to mode-mode
couplings.

We conclude this section by commenting on what is achieved by going
beyond the CBF approximation. In that treatment, the energy denominator in
the self-energy \eqref{eq:sigma} contains only the Feynman spectrum
$\varepsilon_{\rm F}(k)$.  As a consequence, ``mode-mode'' couplings
would describe only the coupling between Feynman phonons. The most
obvious consequence of this restriction is that the so-called 
``Pitaevskii-plateau,'' caused by the fact that it is kinematically 
permitted for a perturbation to decay into two rotons, appears at twice 
the roton energy of the Feynman spectrum. A less obvious consequence 
is that the area above the zero-sound spectrum is filled by a continuum.

\section{CONCLUSIONS}
\label{sec:conclusions}

We have in this work examined the properties of a fictitious system of
$\alpha$ particles interacting via a local two-body interaction based
on scattering data as well as two versions of the $\alpha-\alpha$
interaction developed from Gogny models of the two-nucleon interaction.  
We have treated the problem as an exercise of modern microscopic quantum 
many-body theory, from which much can be learned with relative ease, 
as in the case of liquid $^4$He, given its boson constituents. In doing 
so, we have stressed the fact that four {\em a-priori\/} rather different 
many-body methods, when developed to a level that they contain the same 
physics, actually lead to the same equations to be solved.  These equations 
are, in fact, quite simple for bosons. The only quantity that must be determined 
by either diagrammatic expansion or phenomenological considerations is 
the ``totally irreducible'' interaction $V_{\rm I}(r)$.  We have chosen 
here the route suggested by Jastrow-Feenberg theory because the 
derivation of the relevant quantities is then by far the simplest.  The 
parquet-diagram summations lead, to the extent that has been determined 
so far \cite{TripletParquet}, to the same answer \cite{MixMonster}.

We have highlighted the similarity of alpha matter to the far better
understood system of liquid $^4$He, but we have also exposed its
significant differences.  From experience with the latter system, we
are confident that practically all of the results presented here are
quantitative, the only exception being those for the condensate
fraction. The phase diagram of $\alpha$ matter should be very similar
to that of \he4 \cite{Wilks} displaying a spinodal decomposition at
low densities, a $\lambda$ transition from a superfluid to a normal liquid
and (with the caveat that model {\em per-se\/} may be invalid at high
densities, a liquid-solid phase transition. In particular there is
no indication that $\alpha$-particles could form a Bose-Einstein
condensate as found in untracold gases.

It is rather straightforward to extend the calculations to $\alpha$
droplets, if that should be of interest, in analogy with the case of
$^4$He droplets \cite{ChinKroPRB,ChinKroPRL,ChinKroCPL,dropdyn}.
However, the far more relevant problem is that of $\alpha$-nucleon
mixtures, a fermionic-bosonic composite, whose treatment could follow
much the same pattern within the Feenberg-Jastrow framework. See
Ref.~\cite{Satarov20} for a mean-field treatment of this problem. 
Variational/parquet theory is available for boson-fermion mixtures
\cite{MixMonster}, as well as for realistic nucleon-nucleon 
interactions \cite{Reid68,Reid93,AV18} \cite{v3eos,v3twist}, 
though necessarily more laborious due to the fermion statistics.


\newpage
\bibliography {papers}
\bibliographystyle{apsrev4-1}
\end{document}